\documentclass[aps,prb,twocolumn,floats,floatfix,showpacs,footinbib]{revtex4}
\pdfoutput=1
\usepackage{calc}
\usepackage{tikz}
\usepackage{graphicx}
\usepackage{color}
\usepackage{amsmath, amssymb}
\usepackage[percent]{overpic}
\usepackage{array}
\usepackage{hyperref}
\usepackage{multirow}
\usepackage{comment}

\newcommand{\avg}[1]{\left< #1 \right>} 

\begin{document}
\title{Finite-size scaling and multifractality at the Anderson transition for the three Wigner-Dyson symmetry classes in three dimensions}
\author{L{\'a}szl{\'o} \surname{Ujfalusi}}
\author{Imre \surname{Varga}}
\email[Contact: ]{ujfalusi@phy.bme.hu, varga@phy.bme.hu}
\affiliation{Elm{\'e}leti Fizika Tansz{\'e}k, Fizikai Int{\'e}zet, 
Budapesti M{\H{u}}szaki {\'e}s Gazdas{\'a}gtudom{\'a}nyi Egyetem, 
H-1521 Budapest, Hungary}
\date{\today}
\begin{abstract}
The disorder induced metal--insulator transition is investigated in a three-dimensional simple cubic lattice and compared for the
presence and absence of time-reversal and spin-rotational symmetry, i.e. in the three conventional symmetry classes. Large scale numerical simulations 
have been performed on systems with linear sizes up to $L=100$ in order to obtain eigenstates at the band center, $E=0$. The multifractal dimensions, exponents 
$D_q$ and $\alpha_q$, have been determined in the range of $-1\leq q\leq 2$. The finite-size scaling of the generalized
multifractal exponents provide the critical exponents for the different symmetry classes in accordance with values known from the 
literature based on high precision transfer matrix techniques. The multifractal exponents of the different symmetry classes provide
further characterization of the Anderson transition, which was missing from the literature so far.
\end{abstract}
\maketitle


\section{Introduction}
The metal-insulator transition (MIT), and disordered systems have been at the forefront of condensed matter research since the middle of the last century~\cite{Anderson58}, 
and yet this topic still has several open questions and is still actively investigated. In the last few years experimental evidence has been obtained about this topic; 
in particular, reporting Anderson localization of ultrasound in disordered elastic networks~\cite{Hu08,Faez09}, light in disordered photonic 
lattices in the transverse direction~\cite{Segev13}, or in an ultracold atomic system in a disordered laser trap~\cite{AspectBouyer12}. 
Richardella {\it et al.}~\cite{Richardella10} examined the MIT in a dilute magnetic semiconductor Ga$_{1-x}$Mn$_x$As, which is a strongly interacting and 
disordered system. They found a clear phase transition together with multifractal fluctuations of the local density of states (LDOS) at the Fermi energy, showing, 
that multifractality 
is a robust and important property of disordered systems. Multifractal properties consistent with the theory of Anderson localization are also found in the ultrasound 
system~\cite{Faez09}. On the theoretical side, we know that disorder plays a crucial role in integer quantum Hall effect~\cite{Aoki81}, and recently 
it was shown that an enhanced correlation of multifractal wave-function densities in disordered systems can increase the superconducting critical 
temperature~\cite{Feigelman10Burmistrov12} or the multifractal fluctuations of the LDOS close to criticality may lead to a new phase due to the presence of
local Kondo effects induced by local pseudogaps at the Fermi energy~\cite{Kettemann12}.
Moreover, Anderson localization has also been reported in the spectrum of the Dirac operator within the lattice model of QCD at high temperatures using spectral 
statistics~\cite{Giordano14}, and multifractal analysis seems to corroborate it, as well~\cite{Giordano142}.

These models show an increased interest in understanding the nature of the Anderson transition in the presence of various global
symmetries. A comprehensive review of the current understanding is given in Ref.~\onlinecite{EversMirlin08}. These symmetry classes have been
introduced first to describe random matrix ensembles, but the naming conventions are the same in the field of disordered systems. The classification 
considers two global symmetries: time-reversal and spin-rotational symmetry. As it turns out, beside these symmetries there are three further symmetry 
classes according to the presence of chiral symmetry, and in addition there are four Bogoliubov-de Gennes classes also, corresponding to particle-hole 
symmetry~\cite{EversMirlin08} prominent in hybrid (superconductor-normal) systems. The effect of symmetry classes at the Anderson transition has 
already been investigated earlier~\cite{Varga99} using spectral statistics, but there is much less work based on the multifractal analysis of the eigenstates, 
and multifractal exponents are known numerically only for the orthogonal class~\cite{Rodriguez11}.

Our goal in this article is to fill in this gap and apply multifractal finite-size scaling (MFSS), developed originally by Rodriguez, Vasquez, R\"{o}mer and 
Slevin~\cite{Rodriguez11}, to the Anderson models in the three conventional Wigner-Dyson (WD) classes. The organization of the article is the following.
In Sec.~\ref{sec:anderson_classes_numerics} we define the model and describe its numerical representation. In Sec.~\ref{sec:MFSS} we briefly describe 
the finite-size scaling analysis of the generalized multifractal exponents of the critical eigenstates, in Sec.~\ref{sec:and_MFSS} we give the results obtained 
for the three universality classes and finally in Sec.~\ref{sec:summary} we summarize our results.

\section{Models and numerical representation}\label{sec:anderson_classes_numerics}
\subsection{The model}
In this article we investigate Anderson models belonging to the three WD classes, without chiral and 
particle-hole symmetry. We investigate the case of diagonal disorder and nearest-neighbor hopping, 
therefore the Hamiltonian reads as
\begin{equation}
\mathcal{H}=\sum_{i\sigma} \varepsilon_i c_{i\sigma}^\dagger c_{i\sigma} 
-\sum_{ij\sigma\sigma'}t_{ij\sigma\sigma'} c_{i\sigma}^\dagger c_{j\sigma'} + h.c.,
\end{equation}
where $i,j$ and $\sigma,\sigma'$ stand for site- and spin index, $\varepsilon_i$-s are random on-site energies, 
which are uniformly distributed over the interval $\left[-\frac{W}{2},\frac{W}{2}\right]$, $W$ acts as disorder
strength. Using a uniform distribution is just a convention, other distributions of disorder, e.g. Gaussian, 
binary, etc. can be used as well.

In the orthogonal class time-reversal and spin-rotational symmetry are preserved. In this case the 
Hamiltonian is invariant under orthogonal transformations -- hence the name --, therefore it is a real 
symmetric matrix. Since spin does not play a role, we consider a spinless Anderson model. 
In the numerical simulations the Hamiltonian is represented by an $N\times N$ real symmetric matrix, 
where $N=L^3$, and $L$ is the linear system size in lattice spacing. The diagonal elements, are are uniformly distributed 
random numbers, the off-diagonal elements are zero, except if $i$ and $j$ are nearest neighbors:
\begin{equation}
H_{ij}^{O}=\begin{cases} 
\varepsilon_i \in U\left[-\frac{W}{2},\frac{W}{2}\right]\text{, if } i=j\\
-1\text{, if $i$ and $j$ are neighboring sites}\\
0\text{, otherwise}\end{cases}
\end{equation}
The energy unit is fixed by setting the hopping elements to $1$. To avoid surface effects, 
we use periodic boundary conditions. However, this case was investigated very carefully by 
Rodriguez {\it et al.}~\cite{Rodriguez11}, we consider this symmetry class to verify our numerical 
method, and to obtain a complete description of all the WD classes.

In the unitary class time-reversal symmetry is broken, which can be realized physically by applying 
a magnetic field. It can be shown, that either spin rotational symmetry is broken or not, the model will 
belong to the unitary class~\cite{EversMirlin08}. The Hamiltonian is invariant under unitary 
transformations therefore it is a complex hermitian matrix. We discuss the case when 
spin-rotational symmetry is present, because this way we can use spinless fermions again, 
which keeps the matrix size $N\times N$. However, one has to store about twice as much data compared with
the orthogonal case, because here every off-diagonal matrix element is a complex number. 
Obviously finding an eigenvalue and an eigenvector takes more time, too.

For the numerical simulations we followed Slevin and Ohtsuki~\cite{SlevinOhtsuki97}. Let us consider 
a magnetic field pointing in the $y$ direction with flux $\Phi$, measured in units of the flux quantum, 
${h/e}$. Its effect can be represented by a unity phase factor, the Peierls substitution for the hopping 
elements of the Hamiltonian matrix. The upper triangular of the Hamiltonian reads as
\begin{equation}
H_{i\leq j}^{U}=\begin{cases} 
\varepsilon_i \in U\left[-\frac{W}{2},\frac{W}{2}\right] \text{, if } i=j\\
-1 \text{, if $i$ and $j$ are neighboring sites}\\
\phantom{-1\text{, }}\text{in the $x$ or $y$ direction}\\
-e^{i2\pi\Phi x} \text{, if $i$ and $j$ are neighboring}\\
\phantom{-1\text{, }} \text{sites in the $z$ direction}\\
0 \text{, otherwise}\end{cases}
\end{equation}
Complex hermiticity sets the off-diagonal elements in the lower triangular part, $j<i$. Periodic boundary 
conditions and flux quantization force a restriction for the magnetic flux namely, that $\Phi\cdot L$ must be 
an integer. In the thermodynamic limit arbitrarily small magnetic field drives the system from the orthogonal 
to the unitary class. However, in a finite system the relationship between the system size, $L$, and the magnetic 
length, $L_H=\frac{1}{\sqrt{2\pi\Phi}}$ matters. In the case of weak magnetic field, $L\ll L_H$, the system 
belongs to the orthogonal class, in the case of strong magnetic field, $L\gg L_H$, it belongs to the unitary 
class. Since we use system sizes that are multiples of 10 lattice spacings, see 
Tab.~\ref{tab:anderson_classes_fss_systemsize}, we chose $\Phi=\frac{1}{5}$. 
This leads to $L_H\approx 0.892$ therefore this choice clearly fulfills the two conditions above.

In the symplectic class time-reversal symmetry is present, and spin-rotational symmetry is broken, 
which describes a system with spin-orbit interaction. In this case the Hamiltonian is invariant under symplectic 
transformations therefore it is a quaternion hermitian matrix. For the numerical simulations we followed Asada, Slevin 
and Ohtsuki~\cite{AsadaSlevinOhtsuki05}. Since in this case we have to deal with the spin index also, the Hamiltonian 
is an $2N\times 2N$ complex hermitian matrix. Diagonal elements corresponding to the $i$th site and hopping 
elements between sites $i$ and $j$ are $2\times 2$ matrices because of the spin indexes, having a form
\begin{equation}
\epsilon_i=\left(\begin{matrix}
\varepsilon_i & 0 \\
         0          & \varepsilon_i
\end{matrix}\right)
\qquad
t_{ij}=\left(\begin{matrix}
e^{i\alpha_{ij}\cos\beta_{ij}} & e^{i\gamma_{ij}\sin\beta_{ij}} \\
-e^{-i\gamma_{ij}\sin\beta_{ij}} & e^{-i\alpha_{ij}\cos\beta_{ij}}
\end{matrix}\right),
\end{equation}
where $\varepsilon_i$ is an uniformly distributed random on-site energy from the interval $\left[-\frac{W}{2},\frac{W}{2}\right]$, 
$\alpha_{ij},\ \beta_{ij}$ and $\gamma_{ij}$ were chosen to form an SU(2)-invariant parametrization, leading to the so-called 
SU(2) model: $\alpha_{ij}$ and $\gamma_{ij}$ are uniform random variables from the interval $[0,2\pi]$, and $\beta$ has 
a probability density function $p(\beta)d\beta=\sin(2\beta)d\beta$ in the range $\left[0,\frac{\pi}{2}\right]$. 
The upper triangular of the Hamiltonian has the following form:
\begin{equation}
H_{i\leq j}^{S}=\begin{cases} 
\epsilon_i\text{, if } i=j\\
t_{ij}\text{, if $i$ and $j$ are neighboring sites}\\
0\text{, otherwise}\end{cases}
\end{equation}
The off-diagonal elements are defined following complex hermiticity. To store the Hamiltonian requires 
about eight times more memory compared to the orthogonal case, because here every off-diagonal element contains 
four complex numbers. Finding an eigenvalue is much slower than for the unitary case, mainly because of the linear 
size of the matrix is twice as large.

\subsection{Numerical method}
MFSS deals with the eigenvectors of the Hamiltonian, which is a large sparse matrix. Recent high precision 
calculations~\cite{Rodriguez11} use Jacobi-Davidson iteration with incomplete LU preconditioning, therefore we 
decided to use this combination. For preconditioning the ILUPACK~\cite{Bollhofer08} was used, 
for the JD iteration the PRIMME~\cite{Stathopoulos10} package was used. Since the metal-insulator transition occurs at the 
band center~\cite{EversMirlin08} ($E=0$) at disorder $W_c^O\approx 16.5$ for the orthogonal, at $W_c^U\approx 18.3$ 
for the unitary (depending on the strength of magnetic field),  at $W_c^S\approx 20$ for the symplectic class (for our 
parameters), most works study the vicinity of these points. To have the best comparison, we analyzed this regime, 
therefore $20$ disorder values were taken from the range $15\leq W \leq 18$ for the orthogonal class, $23$ 
disorder values were taken from the interval $17\leq W \leq 20$ for the unitary class, and $20$ disorder values 
were taken from the interval $19.4\leq W \leq 20.5$ for the symplectic class. System sizes were taken from the 
range $L=20..100$, and the number of samples are listed in Tab.~\ref{tab:anderson_classes_fss_systemsize}.
\begin{table}
	\begin {center}
	\begin{tabular}{|c| c|}
	\hline
	system size $(L)$ & number of samples\\ \hline
	20 & 15000 \\ \hline
	30 & 15000 \\ \hline
	40 & 15000 \\ \hline
	50 & 15000 \\ \hline
	60 & 10000\\ \hline
	70 & 7500\\ \hline
	80 & 5000\\ \hline
	90 & 4000\\ \hline
	100 & 3500\\ \hline
	\end{tabular}
	\caption{System sizes and number of samples for the simulation for each WD symmetry class.}
	\label{tab:anderson_classes_fss_systemsize}
	\end{center}
\end{table}
We considered only one wave-function per realization, the one with energy closest to zero in order to avoid 
correlations between wave-functions of the same system~\cite{Rodriguez11}.

\section{Finite size scaling laws for generalized multifractal exponents}
\label{sec:MFSS}
In recent high--precision calculations~\cite{Rodriguez11} the multifractal exponents (MFEs) of the eigenfunctions 
of the Hamiltonian have been used to describe the Anderson metal--insulator transition. We use almost the same notation 
and methods as Ref.~\onlinecite{Rodriguez11}, but for better understanding here we introduce shortly the most important quantities 
and notations. The method has recently been successfully extended for the investigation of the quantum percolation transition in three
dimensions~\cite{UjfalusiPRB14}. 

Considering a $d$-dimensional cubic lattice with linear size $L$, one can divide this lattice into smaller boxes with linear size 
$\ell$. If $\Psi$ is an eigenfunction of the Hamiltonian, the probability corresponding to the $k$th box reads as
\begin{equation}  \label{eq:multifractal_mu}
\mu_k=\sum_{i\in box_k} |\Psi_i|^2.
\end{equation}
One can introduce the $q$th moment of the box probability (frequently called generalized inverse participation ratio, GIPR), 
and its derivative:
\begin{equation}\label{eq:multifractals_SqRq} 
R_q=\sum_{k=1}^{\lambda^{-d}} \mu_k^q\qquad 
S_q=\frac{dR_q}{dq}=\sum_{k=1}^{\lambda^{-d}} \mu_k^q \ln \mu_k.
\end{equation}
The average of $R_q$ and $S_q$ follows a power-law behavior as a function of $\lambda=\frac{\ell}{L}$, with exponent $\tau_q$ and $\alpha_q$:
\begin{equation} 
\tau_q=\lim_{\lambda\to 0}\frac{\ln \avg{R_q}}{\ln \lambda}\qquad \alpha_q=\frac{d\tau_q}{dq}=\lim_{\lambda\to 0}\frac{\avg{S_q}}{\avg{R_q}\ln\lambda}.
\label{eq:multifractals_tau_alpha}\end{equation}
$\tau_q$ can be rewritten in the following form:
\begin{equation} 
\tau_q=D_q(q-1)=d(q-1)+\Delta_q,
\label{eq:multifractals_tau_D_Delta}\end{equation}
where $D_q$ is the generalized fractal dimension, and $\Delta_q$ is the anomalous scaling exponent. Employing a Legendre-transform 
on $\tau_q$, we obtain the singularity spectrum, $f(\alpha)$:
\begin{equation} 
    f(\alpha_q)=q\alpha_q-\tau_q.
\label{eq:multifractal_falpha_tauq}
\end{equation}
$\tau_q$, $\alpha_q$, $D_q$ and $\Delta_q$ are often referred to as multifractal exponents. 

According to recent results~\cite{Mirlin06} a symmetry relation exists for $\alpha_q$ and $\Delta_q$ given in the form:
\begin{equation} 
    \Delta_q=\Delta_{1-q}\qquad\qquad \alpha_q+\alpha_{1-q}=2d 
\label{eq:multifractals_Deltaalphasymmety}
\end{equation}
For numerical approaches one has to define the finite-size version of these MFEs at a particular value of disorder:
\begin{eqnarray} 
\tilde{\alpha}_q^{ens}(W,L,\ell) &=& \frac{\avg{S_q}}{\avg{R_q}\ln\lambda}\\
\tilde{D}_q^{ens}(W,L,\ell) &=& \frac{1}{q-1}\frac{\ln\avg{R_q}}{\ln\lambda},
\end{eqnarray} 
where {\it ens} stands for {\it ensemble} averaging over the different disorder realizations. One may define {\it typical} averaged versions also:
\begin{eqnarray} 
\tilde{\alpha}_q^{typ}(W,L,\ell) &=& \avg{\frac{S_q}{R_q}}\frac{1}{\ln\lambda}\\
\tilde{D}_q^{typ}(W,L,\ell) &=& \frac{1}{q-1}\frac{\avg{\ln R_q}}{\ln\lambda}.
\end{eqnarray} 
Similarly to $\tilde{\alpha}_q$ and $\tilde{D}_q$, $\tilde{\Delta}_q$ or $\tilde{\tau}_q$ can be defined, which are 
called generalized multifractal exponents (GMFEs). Every GMFE approaches the value of the corresponding MFE at the critical point, $W=W_c$, 
only in the limit $\lambda\to 0$. We would like to emphasize, that MFEs are defined through {\it ensemble} averaging in principle 
(see Eq.~(\ref{eq:multifractals_tau_alpha})), and {\it ensemble} and {\it typical} averaged MFEs are equal only in a range of $q$, 
$q_{-}<q<q_{+}$~\cite{EversMirlin08}, defined by the two zeros of the singularity spectrum, $f(\alpha_{q_-})=f(\alpha_{q_+})=0$. 
Therefore when in Sec.~\ref{sec:and_MFSS_vary_lambda} we compute MFEs, we will use {\it ensemble} averaged quantities only.

The choice of the investigated range of $q$ is influenced by the following three effects. If $q$ is large, the $q$th power in Eq.~(\ref{eq:multifractals_SqRq}) 
enhances the numerical and statistical errors, leading to a noisy dataset. If $q$ is negative with large absolute value, 
the relatively less precise small wave-function values dominate the sums in Eq.~(\ref{eq:multifractals_SqRq}), which also 
results in a noisy dataset. These two effects together lead to a regime $q_{min}\leq q \leq q_{max}$, where GMFEs behave numerically the best. 
The third effect is coarse graining which suppresses the noise. For $\ell>1$ in an $\ell\times\ell\times\ell$ sized box positive and negative errors 
on the wave-functions can cancel each other. Moreover, in a box large and small wave-function amplitudes appear together with high probability, 
and this way the relative error of a $\mu_k$ box probability is reduced. In other words coarse graining has a nice smoothing effect, which can 
help to widen the range of $q$ that can be investigated.

The renormalization flow of the AMIT has three fixed points: a metallic, an insulating and a critical one. In the metallic fixed point every 
state is extended with probability one therefore the effective size of the states grows proportional to the volume, leading to $D_q^{met}\equiv d$. 
In the insulating fixed point every state is exponentially localized, the effective size of a state does not change with changing system size, 
resulting in $D_q^{ins}\equiv 0$ for $q> 0$, and $D_q^{ins}\equiv \infty$ for $q< 0$. Renormalization does not change the system at criticality, 
therefore it is scale independent, which means self-similarity. 
Therefore wave-functions are supposed to be multifractals, in other words generalized fractals~\cite{janssen}.

Close to the critical point due to standard finite-size scaling arguments one can derive the following scaling laws for the exponents 
$\tilde{\alpha}_q$ and $\tilde{D}_q$ defined above as:
\begin{subequations}
\begin{align}
\tilde{\alpha}_q(W,L,\ell) &= \alpha_q+\frac{1}{\ln\lambda}\mathcal{A}_q\left(\frac{L}{\xi},\frac{\ell}{\xi}\right)
\label{eq:fss_alphaWLl}\\
\tilde{D}_q(W,L,\ell) &= D_q+\frac{q}{\ln\lambda}\mathcal{T}_q\left(\frac{L}{\xi},\frac{\ell}{\xi}\right)
\label{eq:fss_DWLl}
\end{align}
\end{subequations}
Equations (\ref{eq:fss_alphaWLl})--(\ref{eq:fss_DWLl}) can be summarized in one equation:
\begin{equation} \tilde{G}_q(W,L,\ell) = G_q+\frac{1}{\ln\lambda}\mathcal{G}_q\left(\frac{L}{\xi},\frac{\ell}{\xi}\right)\label{eq:fss_scalinglaw_Ll}\end{equation}
$(L,\ell)$ on the left-hand side and $\left(\frac{L}{\xi},\frac{\ell}{\xi}\right)$ on the right-hand side can be changed to $(L,\lambda)$ and $\left(\frac{L}{\xi},\lambda\right)$:
\begin{equation} \tilde{G}_q(W,L,\lambda) = G_q+\frac{1}{\ln\lambda}\mathcal{G}_q\left(\frac{L}{\xi},\lambda\right)\label{eq:fss_scalinglaw_Llambda}
\end{equation}
Our central goal is to fit the above formulas to the numerically obtained data, where $W_c,\ \nu,\ y$ and $G_q$ appear 
among the fit parameters. This fit procedure will provide us the physically interesting quantities and their confidence intervals. 
In the next sections we present different methods for the finite-size scaling.

\subsection{finite-size scaling at fixed $\lambda$}
\label{sec:fss_fixed_lambda}
At fixed $\lambda$, $G_q$ in Eq.~(\ref{eq:fss_scalinglaw_Llambda}) can be considered as the constant term of $\mathcal{G}_q$, 
therefore
\begin{equation} 
     \tilde{G}_q(W,L) = \mathcal{G}_q\left(\frac{L}{\xi}\right),
\label{eq:fss_anderson_scalinglaw}
\end{equation}
where the constant $\lambda$ has been dropped. $\mathcal{G}_q$ can be expanded with one relevant, $\varrho(w)$, 
and one irrelevant operator, $\eta(w)$, the following way by using $w=W-W_c$:
\begin{equation} 
     \mathcal{G}_q\left(\varrho L^{\frac{1}{\nu}}, \eta L^{-y}\right) = 
\mathcal{G}^{r}_q\left(\varrho L^{\frac{1}{\nu}}\right) + \eta L^{-y}\mathcal{G}^{ir}_q\left(\varrho L^{\frac{1}{\nu}}\right)
\label{eq:fss_anderson_scalinglaw_lambda}
\end{equation}
All the disorder-dependent quantities in the above formula can be expanded in Taylor-series:
\begin{eqnarray} 
     \mathcal{G}^{r}_q\left(\varrho L^{\frac{1}{\nu}}\right)&=&\sum_{i=0}^{n_{r}} a_i\left(\varrho L^{\frac{1}{\nu}}\right)^i\\
     \mathcal{G}^{ir}_q\left(\varrho L^{\frac{1}{\nu}}\right)&=&\sum_{i=0}^{n_{ir}} b_i\left(\varrho L^{\frac{1}{\nu}}\right)^i\\
     \varrho(w)=w+\sum_{i=2}^{n_{\varrho}}c_i w^i &\quad& \eta(w)=1+\sum_{i=1}^{n_{\eta}}d_i w^i
\end{eqnarray}
The advantage of this method is, that in the Taylor-series only one variable appears, $\varrho L^{\frac{1}{\nu}}$, 
therefore the number of parameters (including $W_c,\nu\text{ and }y$) is $n_{r}+n_{ir}+n_{\rho}+n_{\eta}+4$, which grows linearly with the expansion orders. 
This method is very effective for computing $W_c,\ \nu$, and $y$, but since $\lambda$ is fixed, one cannot obtain the MFEs. 
In all cases we used $\lambda=0.1$, because it leads to excellent results in Ref.~\onlinecite{Rodriguez11}. It seems, that it is small enough to capture the 
details of a wave-function, and it allows many different system sizes in the range of 
$20\leq L \leq 100$, which we investigated. This way we can also compare our results to those of Ref.~\onlinecite{Rodriguez11} very well.

\subsection{Finite size scaling for varying $\lambda$}
\label{sec:fss_varying_lambda}
In order to take into account different values of $\lambda$ the scaling law given in Eq.~(\ref{eq:fss_scalinglaw_Ll}) 
has to be considered. The expansion of $\mathcal{G}$ in (\ref{eq:fss_scalinglaw_Ll}) is
\begin{eqnarray*} & &\mathcal{G}_q\left(\varrho L^{\frac{1}{\nu}},\varrho \ell^{\frac{1}{\nu}}, \eta' L^{-y'}, 
\eta \ell^{-y}\right) = \mathcal{G}^{r}_q\left(\varrho L^{\frac{1}{\nu}}, \varrho \ell^{\frac{1}{\nu}}\right) +\\
 & &+\eta' L^{-y'}\mathcal{G'}^{ir}_q\left(\varrho L^{\frac{1}{\nu}},\varrho \ell^{\frac{1}{\nu}}\right) +
\eta \ell^{-y}\mathcal{G}^{ir}_q\left(\varrho L^{\frac{1}{\nu}},\varrho \ell^{\frac{1}{\nu}}\right).
\label{eq:fss_MFSS_calG}
\end{eqnarray*}
According to Rodriguez {\it et al.}~\cite{Rodriguez11} the most important irrelevant term is the one containing the finite 
box size, $\ell$, therefore we took into account that one only. This leads to
\begin{eqnarray} 
\tilde{G}_q(W,L,\ell) =
G_q &+& \frac{1}{\ln\lambda} \left(\mathcal{G}^{r}_q\left(\varrho L^{\frac{1}{\nu}}, \varrho \ell^{\frac{1}{\nu}}\right)+\right. \nonumber\\
        &+&\left. \eta \ell^{-y}\mathcal{G}^{ir}_q\left(\varrho L^{\frac{1}{\nu}},\varrho \ell^{\frac{1}{\nu}}\right)\right).
\label{eq:fss_anderson_scalinglaw_Ll}
\end{eqnarray}
The Taylor expansions of the above functions are
\begin{eqnarray} 
\mathcal{G}^{r}_q\left(\varrho L^{\frac{1}{\nu}},\varrho \ell^{\frac{1}{\nu}} \right) &=& \sum_{i=0}^{n_{r}}\sum_{j=0}^{i} 
a_{ij}\varrho^i L^{\frac{j}{\nu}} \ell^{\frac{i-j}{\nu}} \\
\mathcal{G}^{ir}_q\left(\varrho L^{\frac{1}{\nu}},\varrho\ell^{\frac{1}{\nu}}\right) 
&=&\sum_{i=0}^{n_{ir}}\sum_{j=0}^{i} b_{ij}\varrho^i L^{\frac{j}{\nu}} \ell^{\frac{i-j}{\nu}} \label{eq:fss_MFSS_Grelirrel}\\
\varrho(w)=w+\sum_{i=2}^{n_{\varrho}}c_i w^i\ & & \ \eta(w)=1+\sum_{i=1}^{n_{\eta}}d_i w^i
\end{eqnarray}
The advantage of this method is, that it provides the MFE, $G_q$, since it is one of the parameters to fit. 
There are many more data to fit compared to the fixed $\lambda$ case. Fixed $\lambda$ means that at a given system 
size one can use GMFEs obtained at a certain value of $\ell$ -- the one that leads to the desired $\lambda$ -- , while in 
this case one can fit to GMFEs obtained at different values of $\ell$. However, these GMFEs are correlated, because 
they are the results of the coarse graining of the same wave-functions with different sizes of boxes. During the fitting 
procedure one has to take into account these correlations, see Sec.~\ref{sec:fss_fit}. Since the relevant and irrelevant 
scaling functions have two variables, $\varrho L^{\frac{1}{\nu}}$ and $\varrho \ell^{\frac{1}{\nu}}$, one has to fit a 
two-variable function with the number of parameters $(n_{r}+1)(n_{r}+2)/2+(n_{ir}+1)(n_{ir}+2)/2+n_{\rho}+n_{\eta}+3$. We can see, 
that the number of parameters grows as $\sim n_{r/ir}^2$, instead of as $\sim n_{r/ir}$ as for fixed $\lambda$. 
This makes the fitting procedure incorporating the correlations definitely much more difficult.

\subsection{General principles for the FSS fit procedures}
\label{sec:fss_fit}
In this section we discuss the details of the methods and criteria we used during the MFSS. In order to fit the scaling 
law Eq.~(\ref{eq:fss_anderson_scalinglaw}) and (\ref{eq:fss_anderson_scalinglaw_Ll}) we used the MINUIT library~\cite{James75}. 
To find the best fit to the data obtained numerically the order of expansion of $\mathcal{G}^{r/ir}_q$, $\varrho$ and $\eta$ 
must be decided by choosing the values of $n_{r}, n_{ir}, n_{\varrho}$ and $n_{\eta}$. Since the relevant operator is more 
important than the irrelevant one we always used $n_{rel}\geq n_{ir}$ and $n_{\varrho}\geq n_{\eta}$. To choose the order 
of the expansion we used basically three criteria. The first criterion we took into account was to check how close the ratio 
$\chi^2/(N_{df}-1)$ approached unity, where $N_{df}$ stands for the number of degrees of freedom. Let us denote the numerically 
obtained data points by $y_i$, the fit function value at the $i$th parameter value by $f_i$, and the correlation matrix 
of the numerically obtained data points by $C$, which can be computed numerically with a similar expression to the variance. 
With these notations $\chi^2$ reads as
\begin{equation}
     \chi^2=\sum_{i,j}(y_i-f_i)\left(C^{-1}\right)_{ij}(y_j-f_j),
\end{equation}
for more details see Ref.~\onlinecite{Rodriguez11}. If the data points are not correlated, $C$ is a diagonal matrix, and 
the expression leads to the usual form:
\begin{equation}
\chi^2=\sum_{i}\frac{(y_i-f_i)^2}{\sigma_i^2}.
\end{equation}
The number of degrees of freedom, $N_{df}$ is the number of data points minus the number of fit parameters. 
A ratio $\chi^2/(N_{df}-1)\approx 1$ means that the deviations from the best fit are of the order of the standard 
deviation (correlation matrix). The second criterion was that the fit has to be stable against changing the expansion orders, 
i.e. adding a few new expansion terms. From the fits that fulfilled the first two criteria we chose the simplest model, 
with the lowest expansion orders. Sometimes we also took into account the error bars, and we chose the model 
with the lowest error bar for the most important quantities ($W_c,\nu$, {\it etc}...), if similar models fulfilled the first two criteria.

The error bars of the best fit parameters were obtained by a Monte-Carlo simulation. The data points are results of averaging 
so due to central limit theorem, they have a Gaussian distribution. Therefore we generated Gaussian random numbers with 
parameters corresponding to the mean of the raw data points and standard deviation (or correlation matrix) of the mean, 
and then found the best fit. Repeating this procedure $N_{\rm MC}=100$ times provided the distribution of the fit parameters. 
We chose $95\%$ confidence level to obtain the error bars. 

\section{Results of the MFSS for the Anderson models in the WD symmetry classes}
\label{sec:and_MFSS}
With the numerical method described in Sec.~\ref{sec:anderson_classes_numerics} we computed an eigenvector for every disorder 
realization of the Hamiltonian. From the eigenvectors every GMFE is computable, for the orthogonal and unitary class the $|\Psi_i|^2$ 
expression in Eq.(\ref{eq:multifractal_mu}) is trivial, and it means summation for the spin-index for the symplectic class, since spatial 
behavior is in our interest. At fixed $q$ exponents $\tau_q$ and $\Delta_q$ are linear transforms of $D_q$, so we used only the 
$\tilde{\alpha_q}$ and $\tilde{D_q}$ GMFEs for the MFSS. We investigate the range $-1\leq q \leq 2$, because GMFEs behave the best 
in this regime for the reasons described in Sec.~\ref{sec:MFSS}.

\subsection{Results of the MFSS at fixed $\lambda=0.1$}
\label{sec:and_MFSS_lambda01}

\begin{figure*}
  \begin {center}
  \begin{tabular}{c c c}
  \begin{overpic}[type=pdf,ext=.pdf,read=.pdf,width=.33\textwidth]{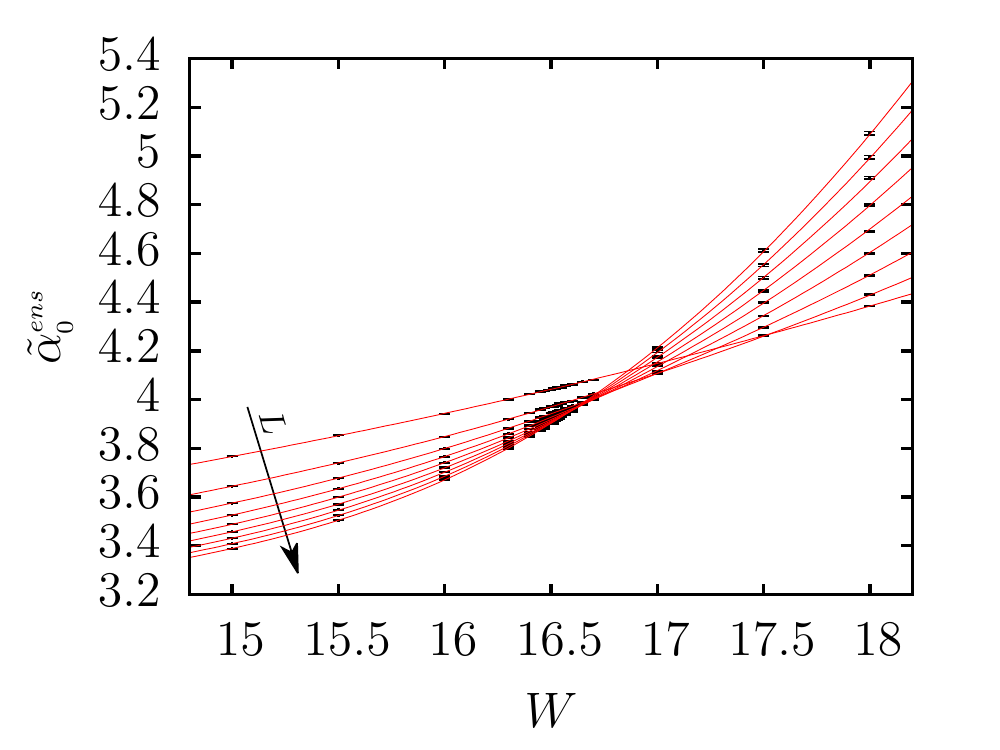}
		\put(20,37){\includegraphics[type=pdf,ext=.pdf,read=.pdf,width=.13\textwidth]{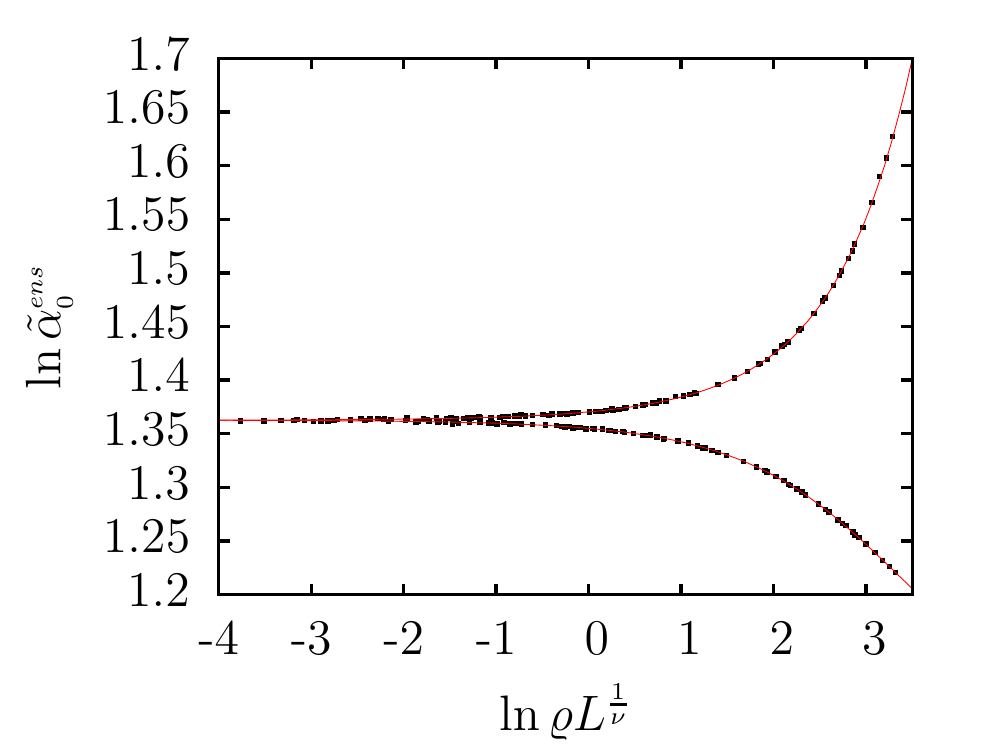}}
		\put(55,25){\scriptsize orthogonal} 
	\end{overpic} &
  \begin{overpic}[type=pdf,ext=.pdf,read=.pdf,width=.33\textwidth]{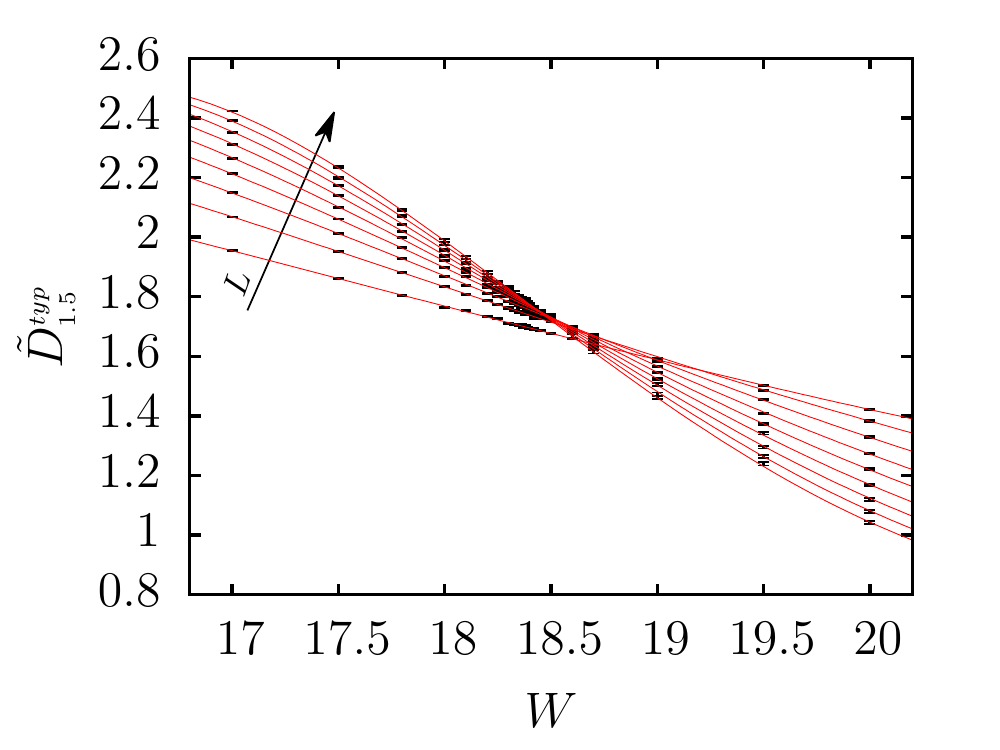} 
		\put(52,39){\includegraphics[type=pdf,ext=.pdf,read=.pdf,width=.13\textwidth]{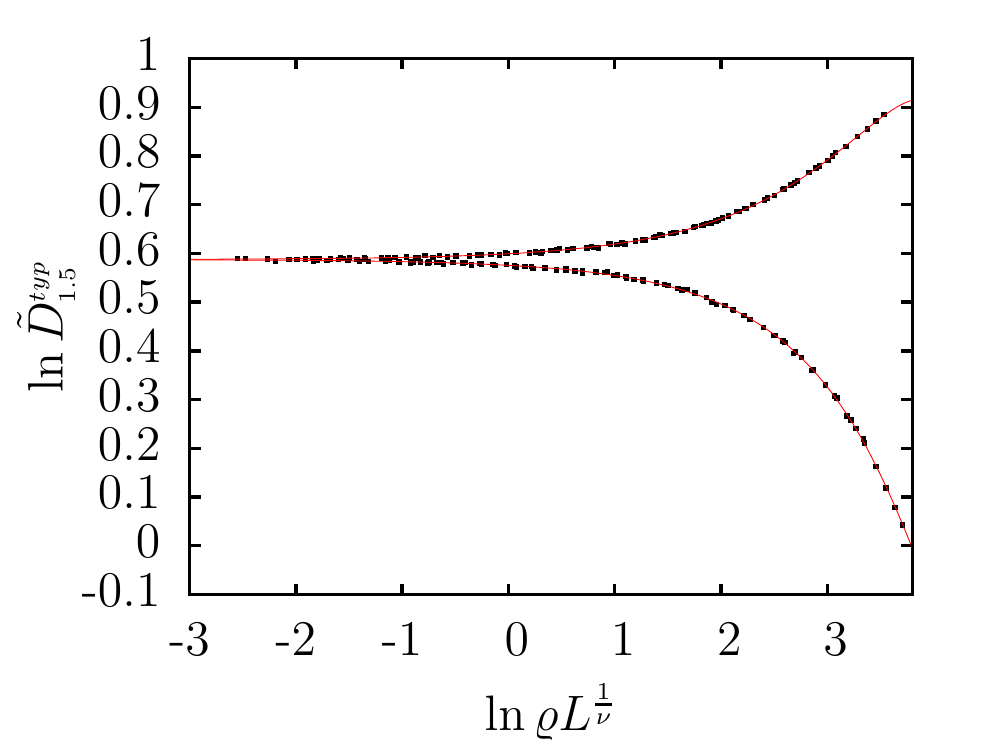}} 
		\put(30,25){\scriptsize unitary}
	\end{overpic}&
  \begin{overpic}[type=pdf,ext=.pdf,read=.pdf,width=.33\textwidth]{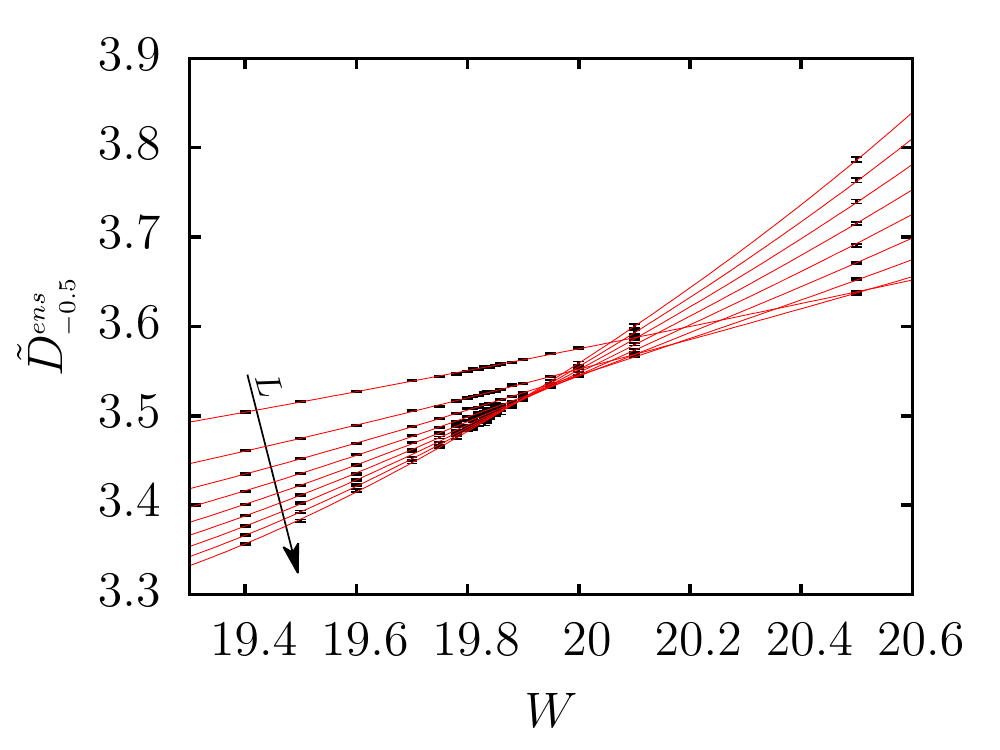} 
		\put(20,38){\includegraphics[type=pdf,ext=.pdf,read=.pdf,width=.13\textwidth]{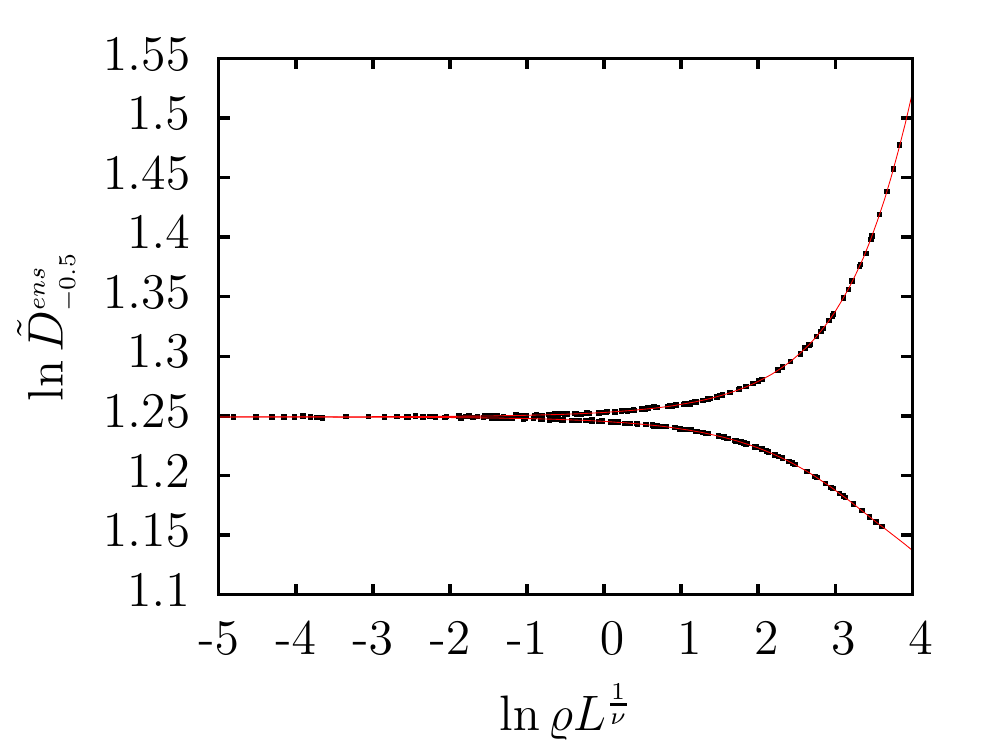}} 
		\put(55,25){\scriptsize symplectic}
	\end{overpic}\\
  \end{tabular}
\caption{Dots are the raw data for different GMFEs in the conventional WD symmetry classes. Red line is the best fit obtained 
by MFSS. Insets are scaling functions on a log-log scale, after the irrelevant term was subtracted. Error bars are shown only on the 
large figures, in order not to overcomplicate the insets.}
  \label{fig:anderson_classes_lambda01_alphaqDq_raw}
  \end{center}
\end{figure*}

The typical behavior of the GMFEs is presented in Fig.~\ref{fig:anderson_classes_lambda01_alphaqDq_raw}. 
In all cases there is a clear sign of phase transition: With increasing system size the GMFEs tend to opposite 
direction on both sides of their crossing point. Note that there is no well-defined crossing point due to the 
irrelevant term in Eq.~(\ref{eq:fss_anderson_scalinglaw_lambda}). Applying the MFSS method described in 
Sec.~\ref{sec:fss_fixed_lambda} with the principles of Sec.~\ref{sec:fss_fit} to the raw data leads to a well-
fitting function, see red lines in Fig.~\ref{fig:anderson_classes_lambda01_alphaqDq_raw}. 
After the subtraction of the irrelevant part from the raw data, plotting it as a function of $\varrho L^{\frac{1}{\nu}}$ 
results a scaling-function also, see insets of Fig.~\ref{fig:anderson_classes_lambda01_alphaqDq_raw}.

\begin{figure*}
  \begin {center}
  \begin{tabular}{c c c}
  \includegraphics[type=pdf,ext=.pdf,read=.pdf,width=.33\textwidth]{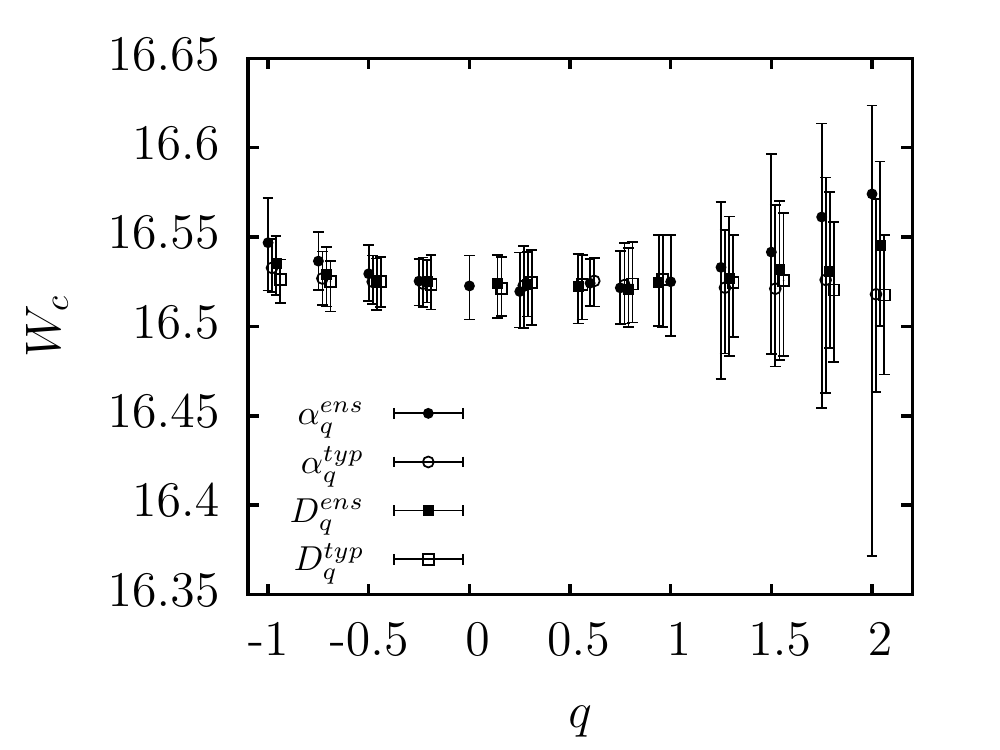} &
  \includegraphics[type=pdf,ext=.pdf,read=.pdf,width=.33\textwidth]{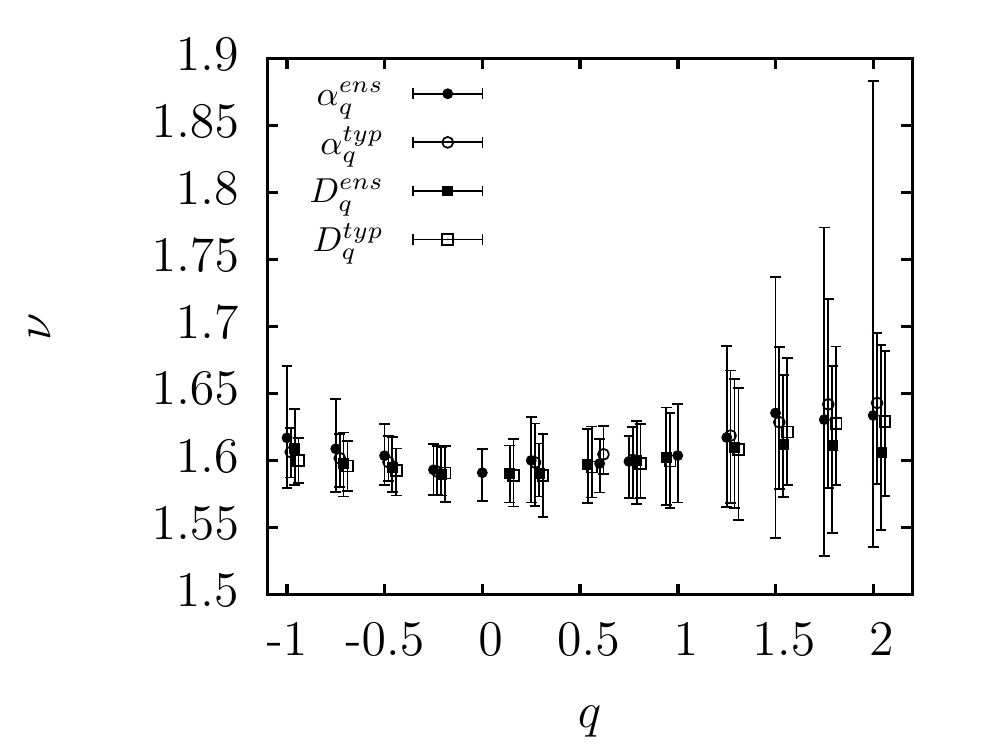} &
  \includegraphics[type=pdf,ext=.pdf,read=.pdf,width=.33\textwidth]{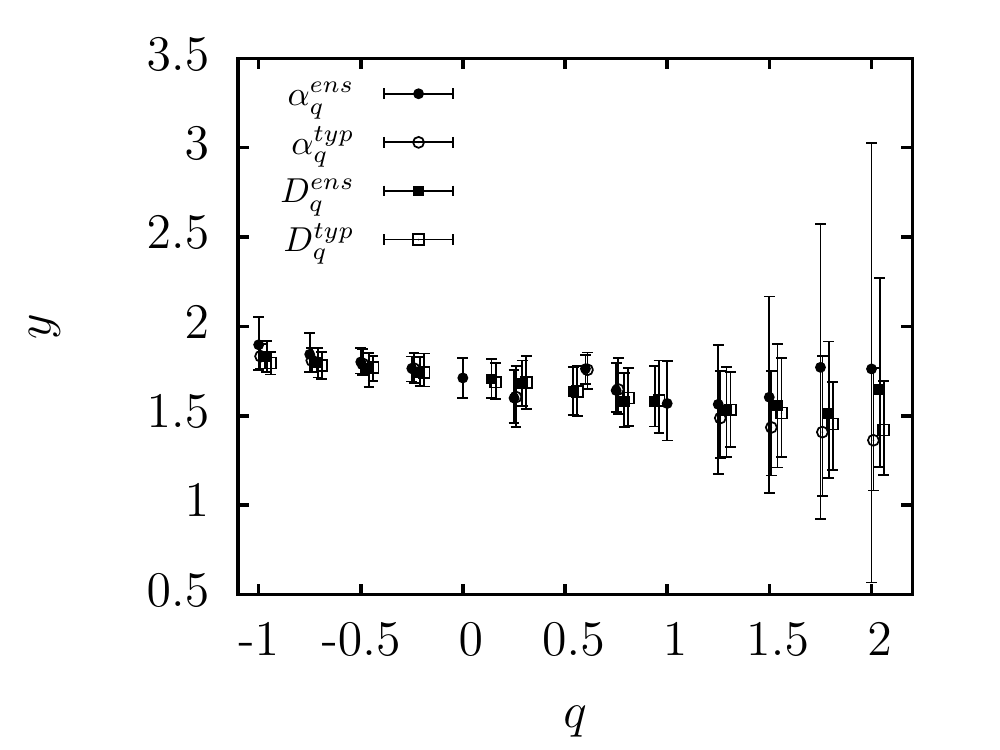} \\
  \includegraphics[type=pdf,ext=.pdf,read=.pdf,width=.33\textwidth]{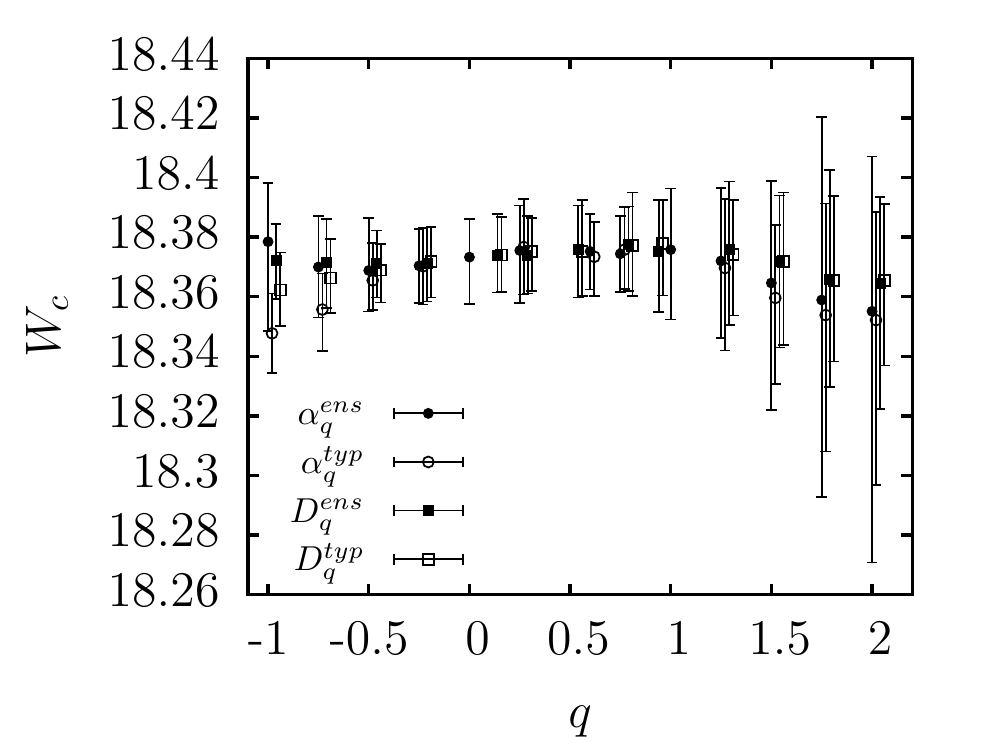} &
  \includegraphics[type=pdf,ext=.pdf,read=.pdf,width=.33\textwidth]{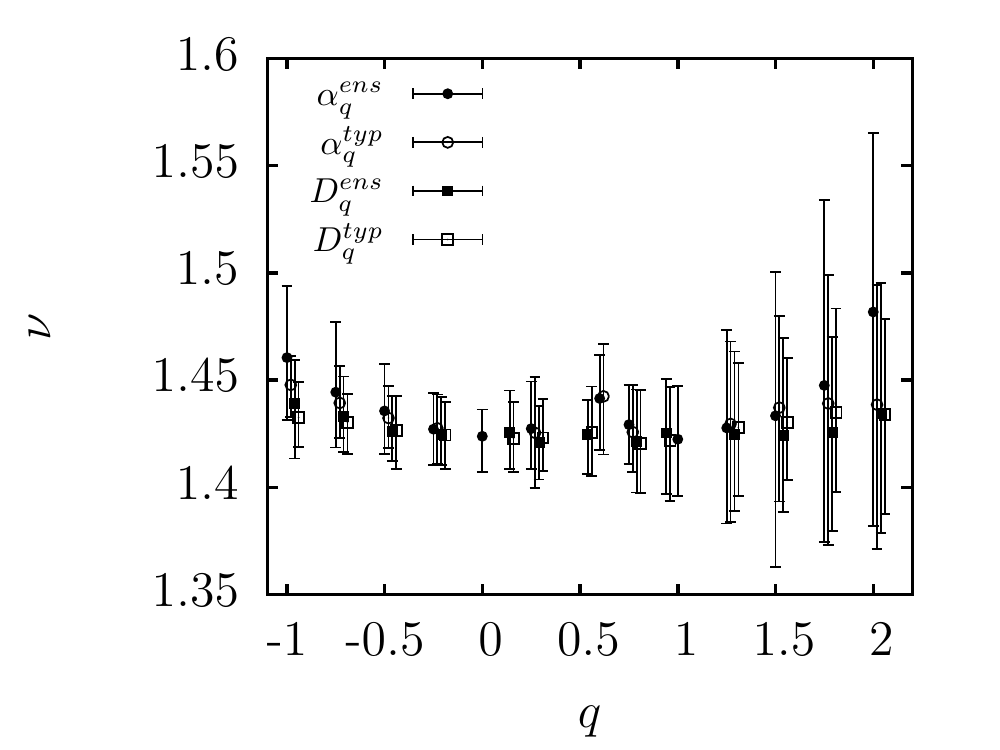} &
  \includegraphics[type=pdf,ext=.pdf,read=.pdf,width=.33\textwidth]{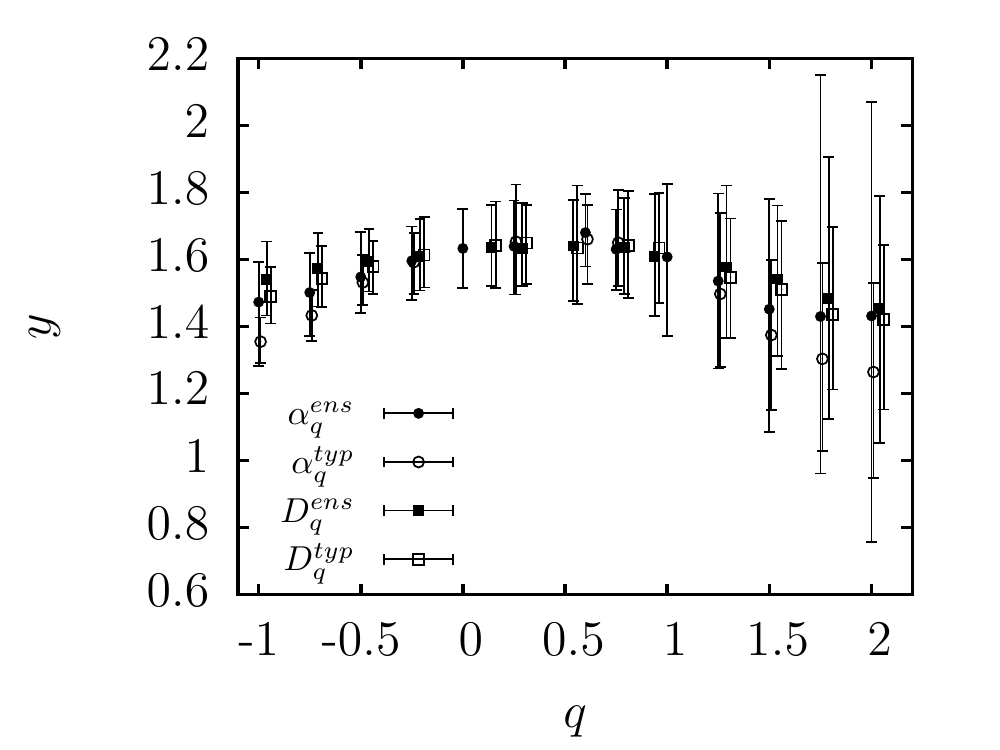} \\
  \includegraphics[type=pdf,ext=.pdf,read=.pdf,width=.33\textwidth]{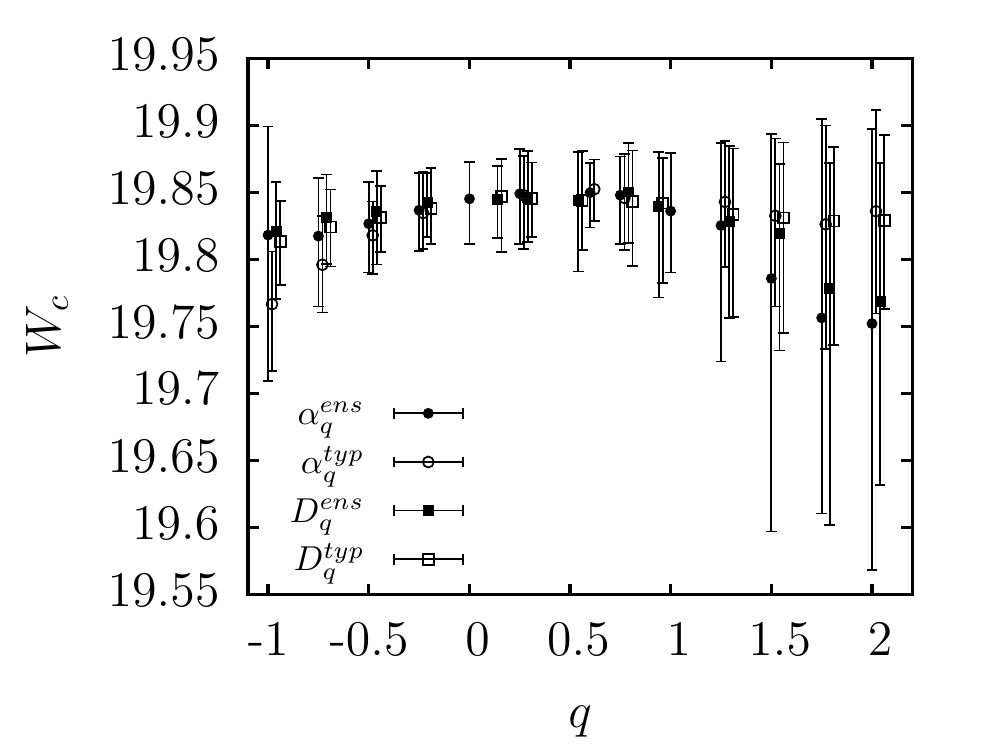} &
  \includegraphics[type=pdf,ext=.pdf,read=.pdf,width=.33\textwidth]{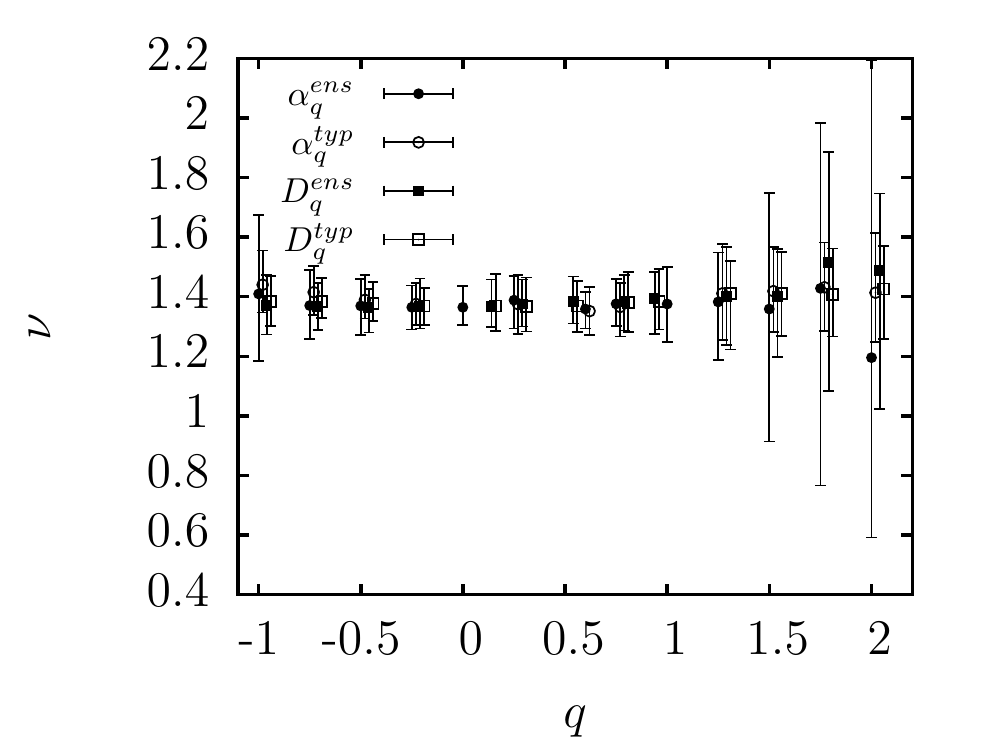} &
  \includegraphics[type=pdf,ext=.pdf,read=.pdf,width=.33\textwidth]{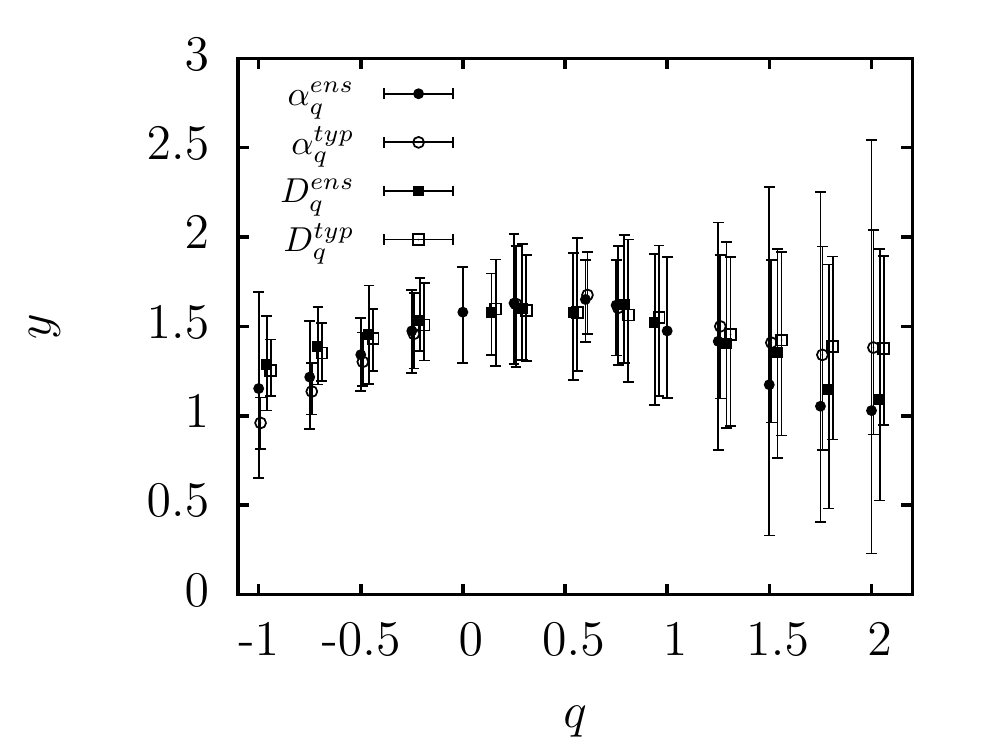} \\
  \end{tabular}
\caption{Critical parameters of the Anderson models in WD classes obtained by MFS at fixed $\lambda=0.1$. 
First row corresponds to the  orthogonal class, second row corresponds to the unitary class, and third row corresponds to the symplectic class.}
  \label{fig:anderson_classes_fitres_lambda01}	
  \end{center}
\end{figure*}

The MFSS provided us the critical point, $W_c$, the critical exponent, $\nu$, and the irrelevant exponent, $y$ at every investigated 
values of $q$, the results are given in Fig.~\ref{fig:anderson_classes_fitres_lambda01}. The parameters of the critical point correspond 
to the system itself, therefore it should not depend on the quantity we used to find it. In other words, it should be independent of $q$, 
the averaging method and the GMFE we used. From Fig.~\ref{fig:anderson_classes_fitres_lambda01} it is clear that this requirement 
is fulfilled very nicely. There is a small deviation for the irrelevant exponent, $y$, obtained from $\alpha^{typ}$ at $q=-1$ and $q=-0.75$ 
in the unitary and symplectic class, but since $y$ describes the subleading part, it is very hard to determine, and we cannot exclude
some sort of underestimatiion of the error bar of this exponent. Another interesting feature of the results is that the error bars get 
larger as $q$ goes above $1$. As written in Sec.~\ref{sec:fss_fit}, large $q$ enhances the errors through the $q$th power in 
Eq.~(\ref{eq:multifractals_SqRq}), leading to bigger error bars. A similar effect can be seen around $q\approx-1$, where the relatively 
less precise small wave-function values dominate the sums in Eq.~(\ref{eq:multifractals_SqRq}), which can also contribute to the deviation 
of $y$ obtained from $\tilde{\alpha}^{typ}$ in this regime. These two effects together lead to our investigated interval $-1\leq q \leq 2$, 
where GMFEs behave the best. The results are strongly correlated, since they were obtained from the same wave-functions, 
therefore they cannot be averaged. We chose a typical $q$-point for every symmetry class to describe the values of the critical parameters, 
see Tab.~\ref{tab:anderson_classes_fitres_lambda01}.

\begin{table*}
	\begin {center}
	\begin{tabular}{|c| c| c| c| c| c| c| c|}
	\hline
class & exp & $W_c^{\lambda}$ & $\nu^{\lambda}$ 
& $y^{\lambda}$ & $N_{df}$ & $\chi^2$ & $n_r n_{ir} n_\varrho n_\eta$\\ \hline
ort & $\tilde{\alpha}_{0.6}^{ens}$ & $16.524\ (16.511..16.538)$ 
& $1.598\ (1.576..1.616)$ & $1.763\ (1.679..1.842)$ & $172$ & $176$ & $3\ 2\ 1\ 0$ \\ \hline
uni & $\tilde{\alpha}_0^{ens/typ}$ & $18.373\ (18.358..18.386)$ 
& $1.424\ (1.407..1.436)$ & $1.633\ (1.516..1.751)$ & $198$ & $179$ & $4\ 2\ 1\ 0$ \\ \hline
sym & $\tilde{D}_{-0.25}^{typ}$ & $19.838\ (19.812..19.869)$ 
& $1.369\ (1.305..1.430)$ & $1.508\ (1.309..1.743)$ & $171$ & $151$ & $4\ 2\ 1\ 0$ \\ \hline
\end{tabular}
\caption{Result of the MFSS at fixed $\lambda=0.1$ for the selected values of $q$.}
\label{tab:anderson_classes_fitres_lambda01}
\end{center}
\end{table*}

In the orthogonal class the critical parameters are in excellent agreement with the most recent high precision results 
of Rodriguez {\it et al.}~\cite{Rodriguez11}, $W_{c\ Rod}^{O\lambda}=16.517\ (16.498..16.533)$, 
$\nu^{O\lambda}_{Rod}=1.612\ (1.593..1.631)$ and $y^{O\lambda}_{Rod}=1.67\ (1.53..1.80)$, 
obtained from $\tilde{\alpha}_0$ with the same method (fixed $\lambda$). This agreement verifies our numerics 
and fit method, and makes it reliable for the other two universality classes.

In the unitary class the critical parameters match with the results of Slevin and Ohtsuki~\cite{SlevinOhtsuki97}, 
$W_{c\ Sle}^{U}=18.375\ (18.358..18.392)$ and $\nu^{U}_{Sle}=1.43\ (1.37..1.49)$, obtained by transfer matrix 
method (they did not published the value of the irrelevant exponent). They used magnetic flux $\Phi=\frac{1}{4}$, 
while we used $\Phi=\frac{1}{5}$, and according to Dr\"{o}se {\it et al.}~\cite{Drose98}, $W_c^U$ depends on the 
applied magnetic flux. However, in Fig.2. of Ref.~\onlinecite{Drose98} it can be seen that the critical points at 
$\Phi=\frac{1}{4}$ and $\Phi=\frac{1}{5}$ are very close to each other, hence the agreement between our critical 
point and the result of Slevin and Ohtsuki.

In the symplectic class the critical parameters agree more or less with the results of Asada 
{\it et al.}~\cite{AsadaSlevinOhtsuki05}, $W_{c\ Asa}^{S}=20.001\ (19.984..20.018)$, $\nu^{S}_{Asa}=1.375\ (1.359..1.391)$ 
and $y^{S}_{Asa}=2.5\ (1.7..3.3)$, obtained by transfer matrix method. However, the difference does not seem to be very large, 
our critical point is considerably different, even though we used exactly the same model. Due to bigger computational 
resources we could investigate much bigger system sizes than they did, therefore it is possible that they underestimated 
the role of the irrelevant scaling, resulting in a somewhat higher critical point.  

The critical points are higher in the unitary and in the symplectic class, than in the orthogonal class, showing that broken 
time-reversal or spin-rotational symmetry requires more disorder to localize wave-functions. Since the value of the critical point 
in the unitary and symplectic class can be influenced by the strength of the applied magnetic flux and spin-orbit coupling, the 
relationship between $W_c^{U\lambda}$ and $W_c^{S\lambda}$ probably depends on these two parameters. However, because 
of their close value of the critical exponents, $\nu^{U\lambda}$ and 
$\nu^{S\lambda}$ are the same within our confidence interval, and the following relation appears: 
$\nu^{O\lambda}>\nu^{U\lambda}\geq\nu^{S\lambda}$. The situation for the irrelevant exponent is similar namely, 
that they are the same within error bar, but $y^{O\lambda}$ seems to be slightly higher than $y^{U\lambda}$, 
which is a bit higher than $y^{S\lambda}$.

\subsection{Results of the MFSS at varying $\lambda$}
\label{sec:and_MFSS_vary_lambda}
As mentioned in Sec.~\ref{sec:fss_fit}, GMFEs obtained by {\it typical} averaging are equal to {\it ensemble}-averaged GMFEs only in a range of $q$, $q_-<q<q_+$. 
Since we intend to compute the MFEs also, we restrict our analysis to {\it ensemble} averaged GMFEs, and drop the label {\it ens} from the notation.

\begin{figure*}
  \begin {center}
  \begin{tabular}{c c c}
  \includegraphics[type=pdf,ext=.pdf,read=.pdf,width=.33\textwidth]{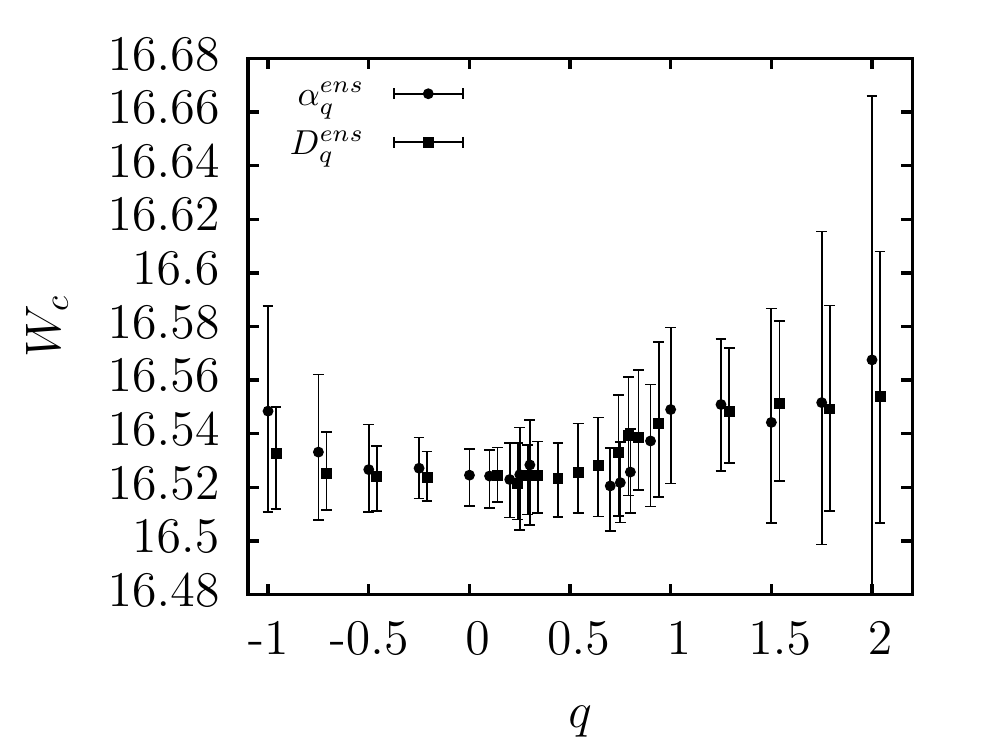} &
  \includegraphics[type=pdf,ext=.pdf,read=.pdf,width=.33\textwidth]{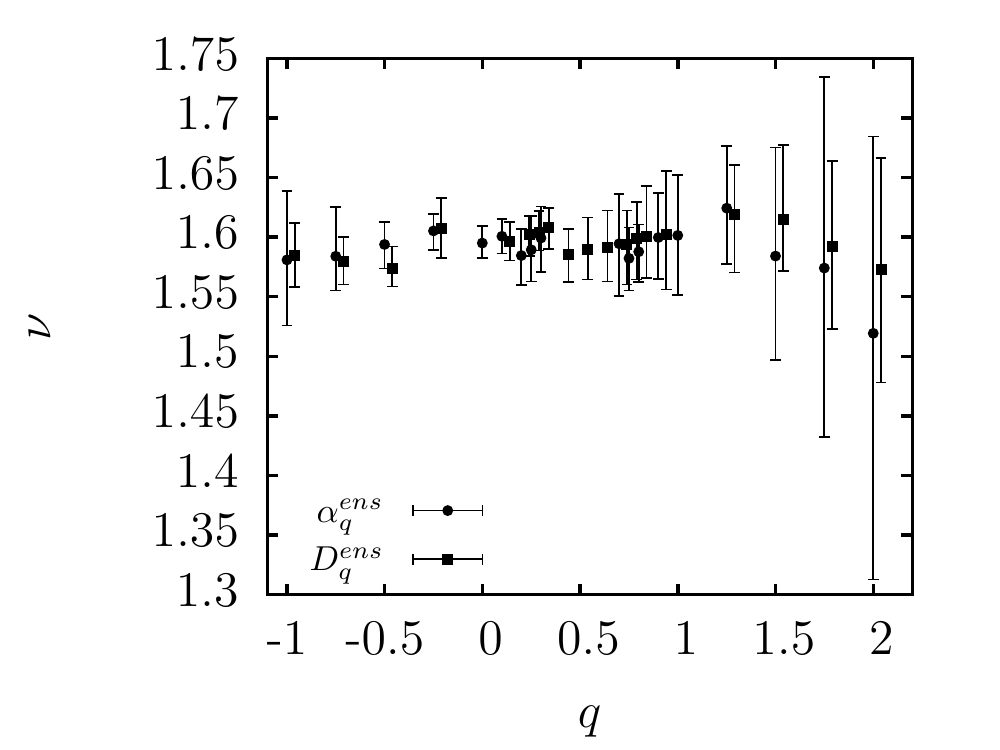} &
  \includegraphics[type=pdf,ext=.pdf,read=.pdf,width=.33\textwidth]{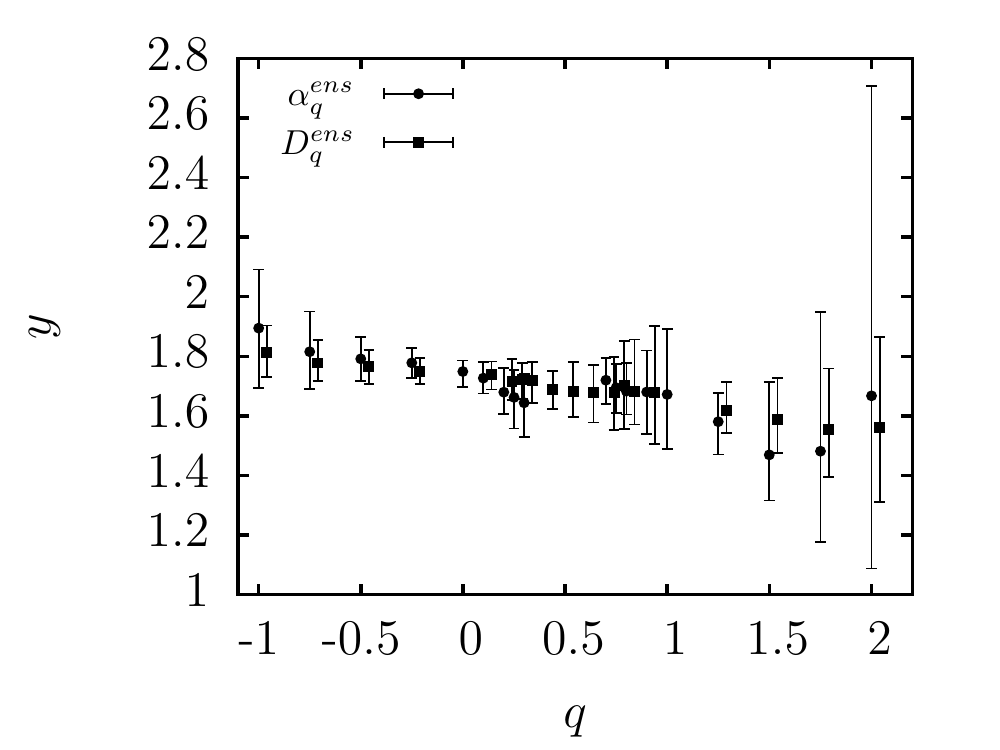} \\
  \includegraphics[type=pdf,ext=.pdf,read=.pdf,width=.33\textwidth]{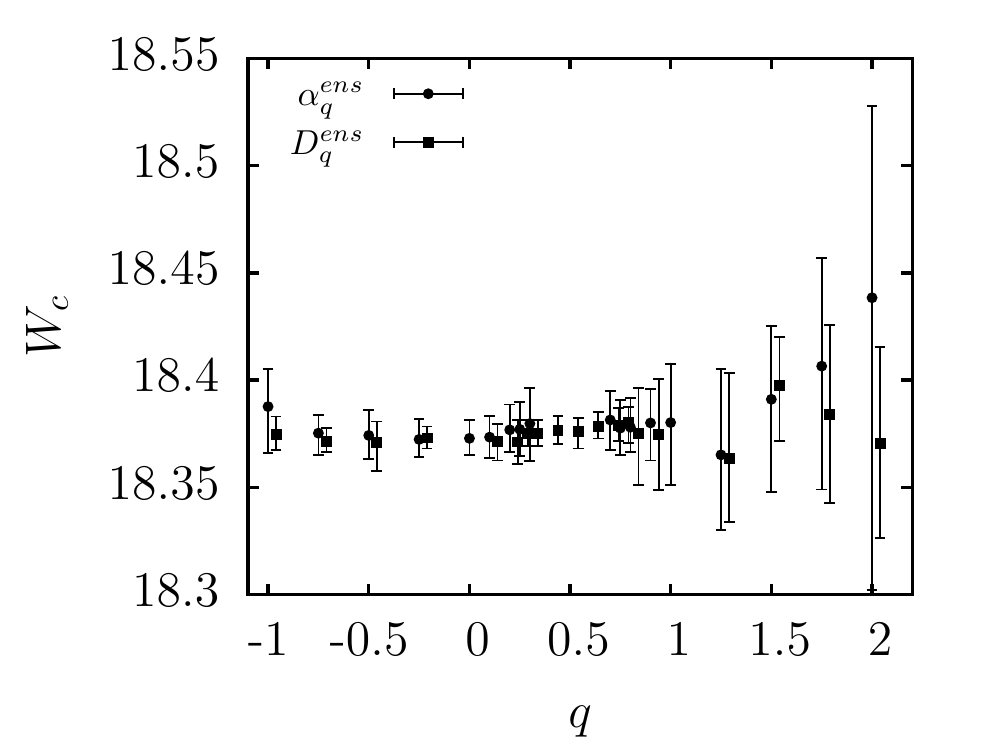} &
  \includegraphics[type=pdf,ext=.pdf,read=.pdf,width=.33\textwidth]{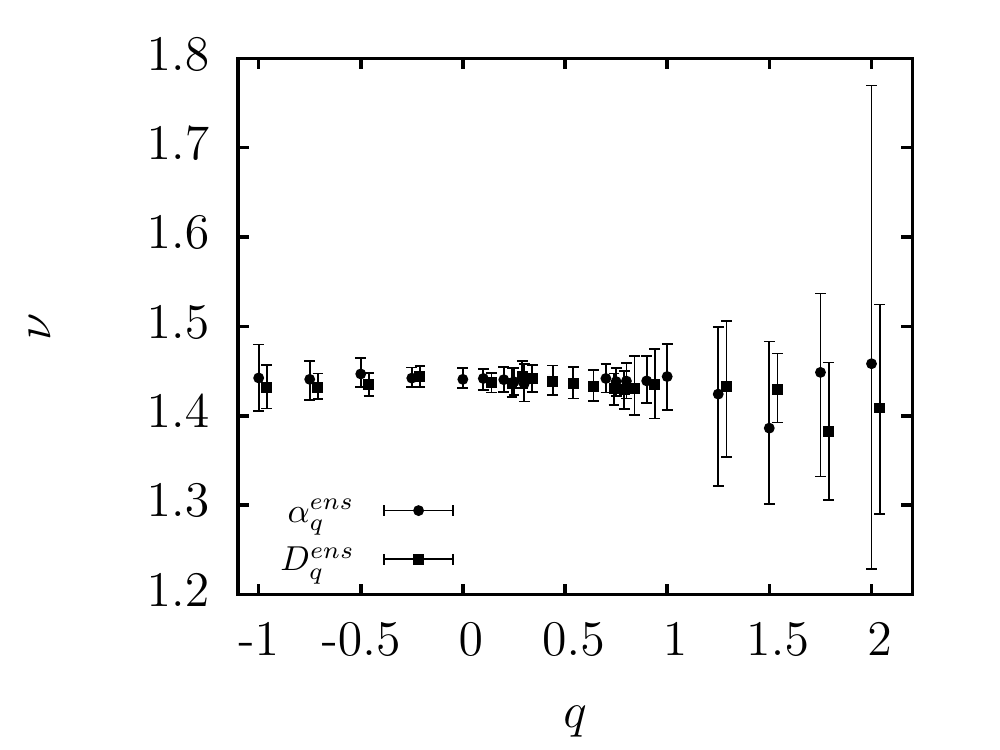} &
  \includegraphics[type=pdf,ext=.pdf,read=.pdf,width=.33\textwidth]{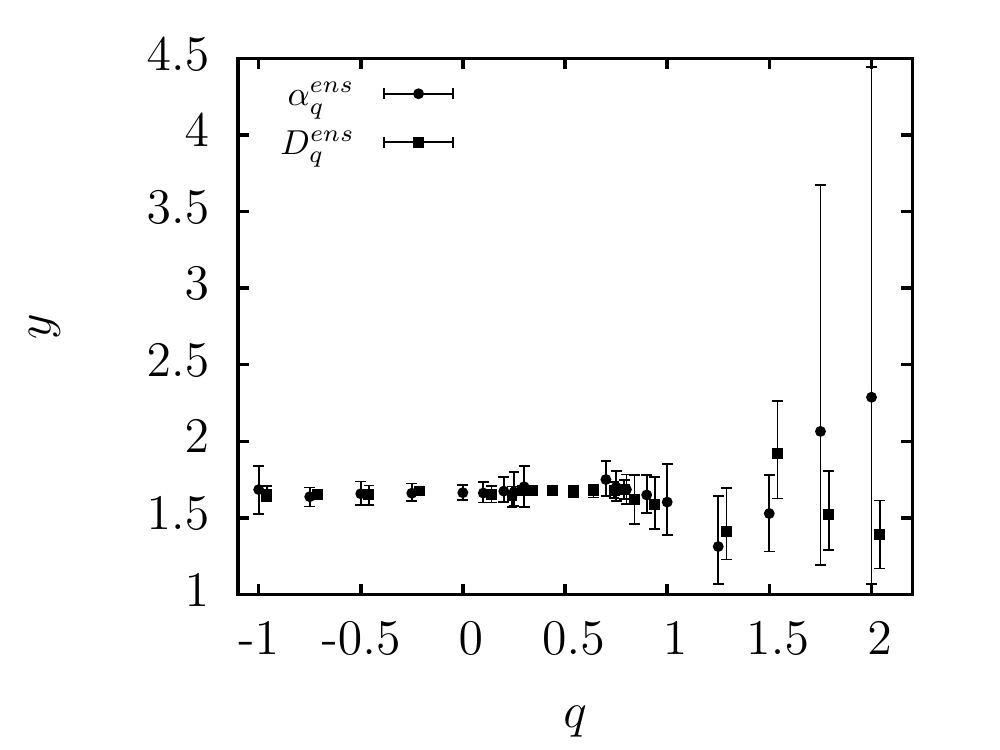} \\
  \includegraphics[type=pdf,ext=.pdf,read=.pdf,width=.33\textwidth]{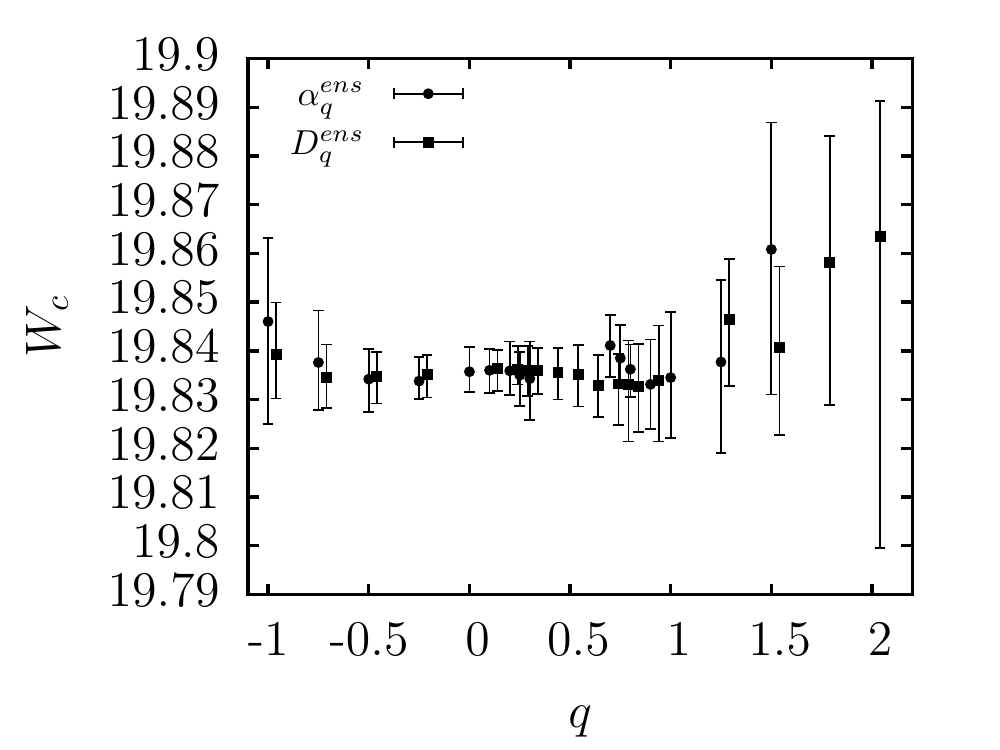} &
  \includegraphics[type=pdf,ext=.pdf,read=.pdf,width=.33\textwidth]{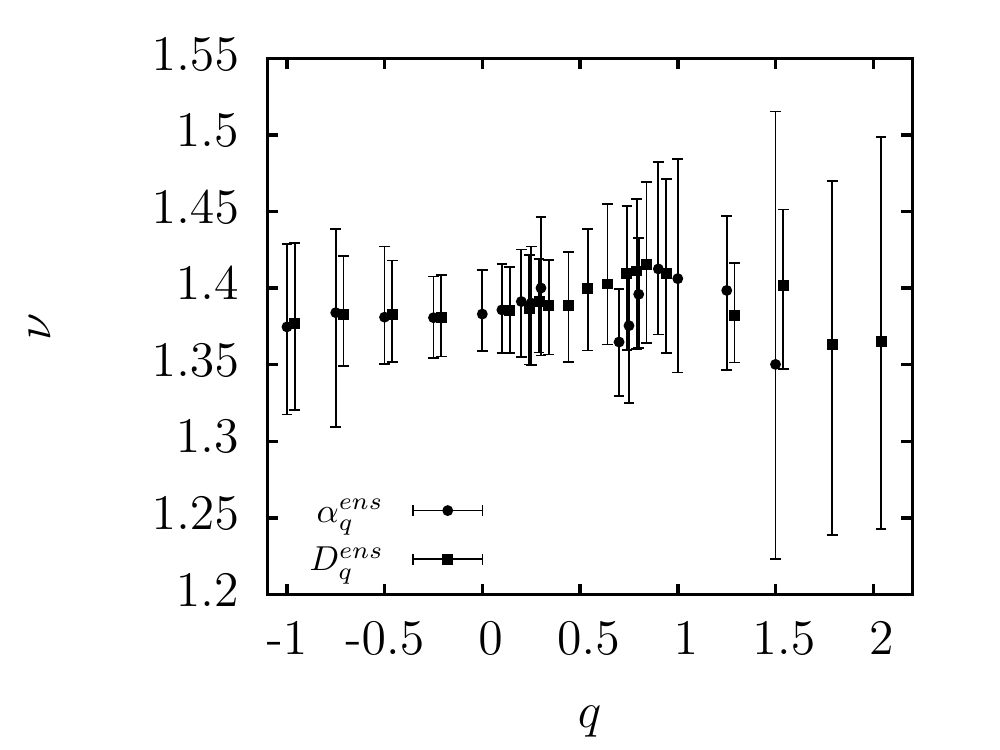} &
  \includegraphics[type=pdf,ext=.pdf,read=.pdf,width=.33\textwidth]{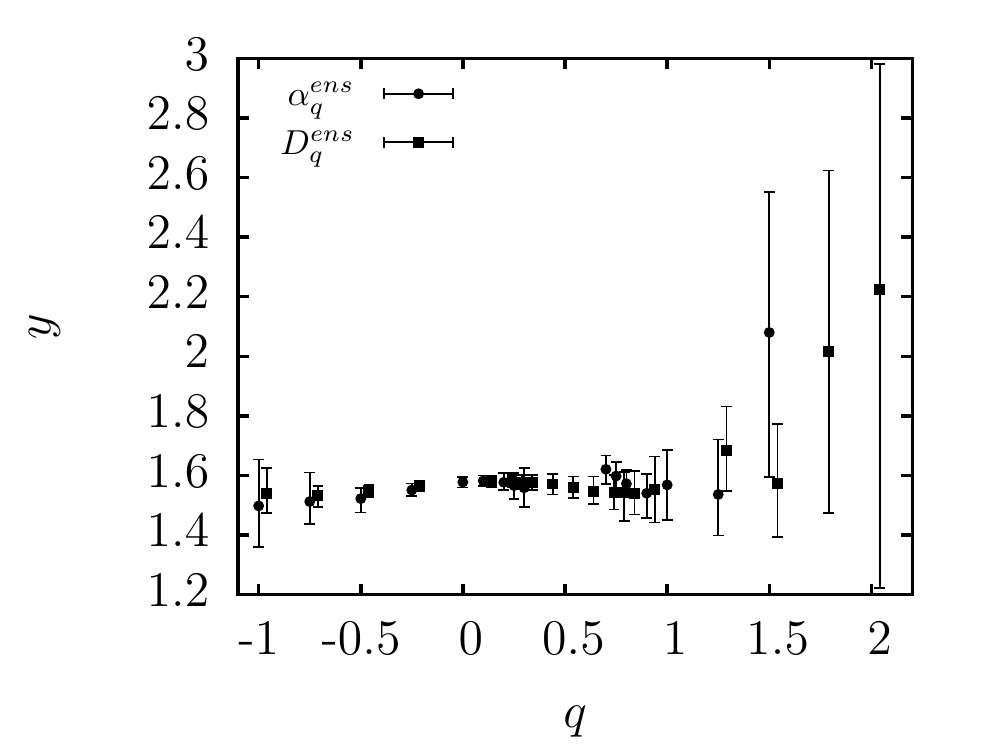} \\
  \end{tabular}
\caption{Critical parameters of the Anderson models in WD classes obtained by two-variable MFFS 
with varying $\lambda$. First row corresponds to the  orthogonal class, second row corresponds 
to the unitary class, and third row corresponds to the symplectic class.} \label{fig:anderson_classes_fitres_vary_lambda}	
  \end{center}
\end{figure*}

We fit the formula Eq.~(\ref{eq:fss_anderson_scalinglaw_Ll})to the raw data. To do that, we choose a range of box size $\ell$, which is used for the MFSS. 
We always use the widest range of $\ell$, that results in convergence, $\chi^2/(N_{df}-1)\approx 1$. We find that for our dataset for different values of $q$ for $\alpha_q$ 
or $D_q$ different ranges of $\ell$ were the best. We used minimal box sizes $\ell_{min}=2$ or $\ell_{min}=3$ and maximal box sizes corresponding to $\lambda_{max}=0.1$ 
or $\lambda_{max}=0.066$. At $\alpha_{0.4}$ and $\alpha_{0.6}$ the fitting method sometimes suffered from convergence troubles and resulted in large error bars, because these 
points are close to the special case of $q=0.5$ where, by definition, $\alpha_{0.5}=d$. Artifacts from this regime were also reported in Ref.~\onlinecite{Rodriguez11}, so 
we decided not to take into account these points for $\alpha$. We tried several combinations of $\ell_{min}$, $\lambda_{max}$ and expansion orders in the {\it symplectic} 
class for $\alpha_{1.75}$ and $\alpha_2$, but none of them resulted in stable fit parameters. Therefore values computed from these points are also missing from our final results, 
which are  visible in Fig.~\ref{fig:anderson_classes_fitres_vary_lambda}. The results are independent of $q$ and the GMFE we used, similar to the fixed $\lambda$ method. In Sec.~\ref{sec:and_MFSS_lambda01} we already saw that according 
to the arguments of Sec.~\ref{sec:fss_fit} error bars get bigger, if $q$ grows beyond $1$. This phenomenon is more amplified here, especially for values coming 
from fits for $\alpha_q$, but larger error bars on values corresponding to $D_q$ are present on a moderate level also. Since Fig.8 of Ref.~\onlinecite{Rodriguez11} shows results 
for this regime only for values corresponding to $\Delta_q$, which is a linear transform of $D_q$, we can compare their results only to ours corresponding to $D_q$. 
One can see that our error bars are similar, even though there are differences probably due to the fact that they used system sizes up to $L=120$, which was not possible for us, 
mainly because of the long runtime and large memory usage for the symplectic model. They also use $\ell_{min}=1$ and $\ell_{min}=2$, while $\ell_{min}=1$ was never suitable 
for our dataset. We do not know the precise origin of this behavior, but we have a few possible explanations. We experience that larger system sizes allow a wider range of $\ell$ 
to be used. We have smaller system sizes than Ref.~\onlinecite{Rodriguez11}, and fewer samples for the largest systems sizes. Noise also gets bigger as $\ell$ decreases,
because of the smoothing effect of boxing described in Sec.~\ref{sec:fss_fit}, which can also explain partly our experience. Another important difference is that 
in Eq.~(37) of Ref.~\onlinecite{Rodriguez11} the authors use an expression in the expansion of the scaling function, which is proportional to the square of the irrelevant term, 
$\left(\eta\ell^{-y}\right)^2$. According to our experience the inclusion of this term produced no improvement in the scaling analysis, so we use the scaling function 
described in Eq.~\ref{eq:fss_anderson_scalinglaw_Ll}. Such a difference might be explained again by our different dataset.

As written in Sec.~\ref{sec:and_MFSS_lambda01}, the results for different values of $q$ are strongly 
correlated, therefore we chose one of them with the lowest error bars that represents well the results 
for that universality class.

The critical parameters listed in Tab.~\ref{tab:anderson_classes_fitres_vary_lambda_chosen} are 
in a very nice agreement with our previous results for the fixed method of $\lambda=0.1$, 
see Sec.~\ref{sec:and_MFSS_lambda01}, and also with the results of 
Refs.~\onlinecite{Rodriguez11,SlevinOhtsuki97,AsadaSlevinOhtsuki05}. Comparing the critical parameters for the 
orthogonal case with the results of Rodriguez {\it et al.}~\cite{Rodriguez11} obtained by the same 
method, $W_{c\ Rod}^{O}=16.530\ (16.524..16.536)$, $\nu_{Rod}^O=1.590\ (1.579..1.602)$, 
we see a nice agreement again. Moreover these results are more accurate with this method 
compared to the fixed $\lambda$ method, leading to (for $y^O$ and $y^U$ only almost) 
significantly different critical exponents and irrelevant exponents for the different WD classes, 
$\nu^O>\nu^U>\nu^S$ and $y^O\geq y^U>y^S$.

\begin{table*}
  \begin {center}
  \begin{tabular}{|c| c| c| c| c| c| c| c|}
  \hline
  class & exp & $W_c$ & $\nu$ & $y$ & $N_{df}$ & $\chi^2$ & $n_r n_{ir} n_\varrho n_\eta$\\ \hline
  ort & $\tilde{\alpha}_0$ & $16.524\ (16.513..16.534)$ & $1.595\ (1.582..1.609)$ & $1.749\ (1.697..1.786)$ & $241$ & $267$ & $3\ 2\ 1\ 0$\\ \hline
  uni & $\tilde{D}_{0.1}$ & $18.371\ (18.363..18.380)$ & $1.437\ (1.426..1.448)$ & $1.651\ (1.601..1.707)$ & $275$ & $232$ & $4\ 2\ 1\ 0$ \\ \hline
  sym & $\tilde{\alpha}_0$ & $19.836\ (19.831..19.841)$ & $1.383\ (1.359..1.412)$ & $1.577\ (1.559..1.595)$ & $361$ & $352$ & $3\ 2\ 1\ 0$ \\ \hline
\end{tabular}
\caption{Critical parameters of the Anderson models in the WD symmetry classes obtained by two-variable 
MFSS with varying $\lambda$.}
\label{tab:anderson_classes_fitres_vary_lambda_chosen}
\end{center}
\end{table*}

\subsection{Analysis of the multifractal exponents}
\begin{figure}
  \begin {center}
  \begin{tabular}{c c}
  \includegraphics[type=pdf,ext=.pdf,read=.pdf,width=.25\textwidth]{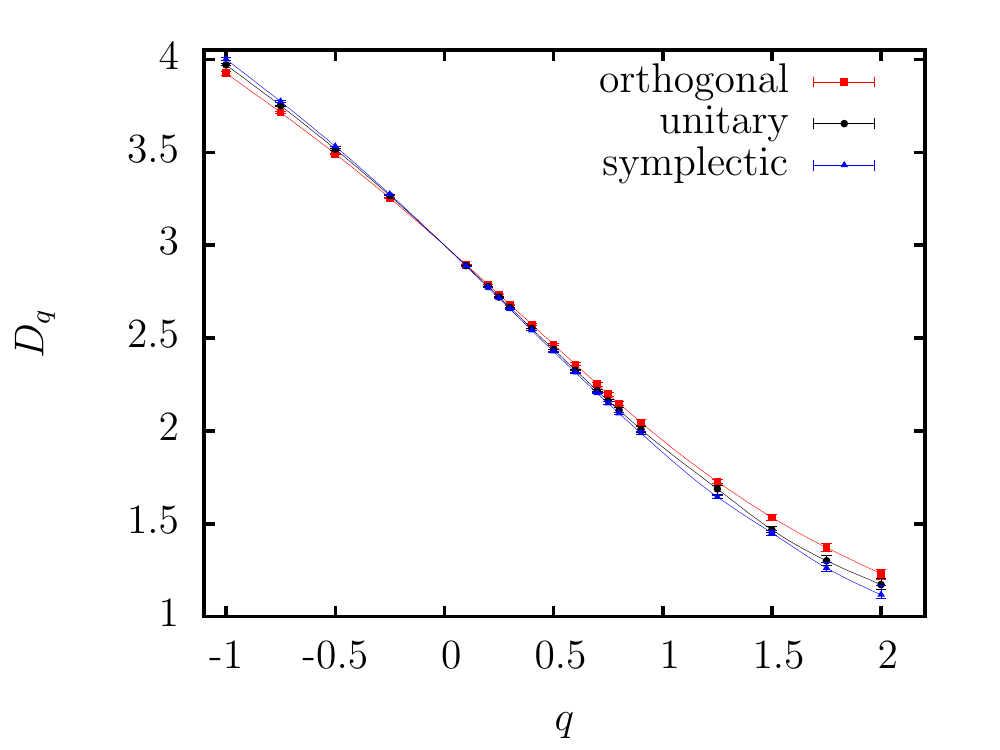} &
  \includegraphics[type=pdf,ext=.pdf,read=.pdf,width=.25\textwidth]{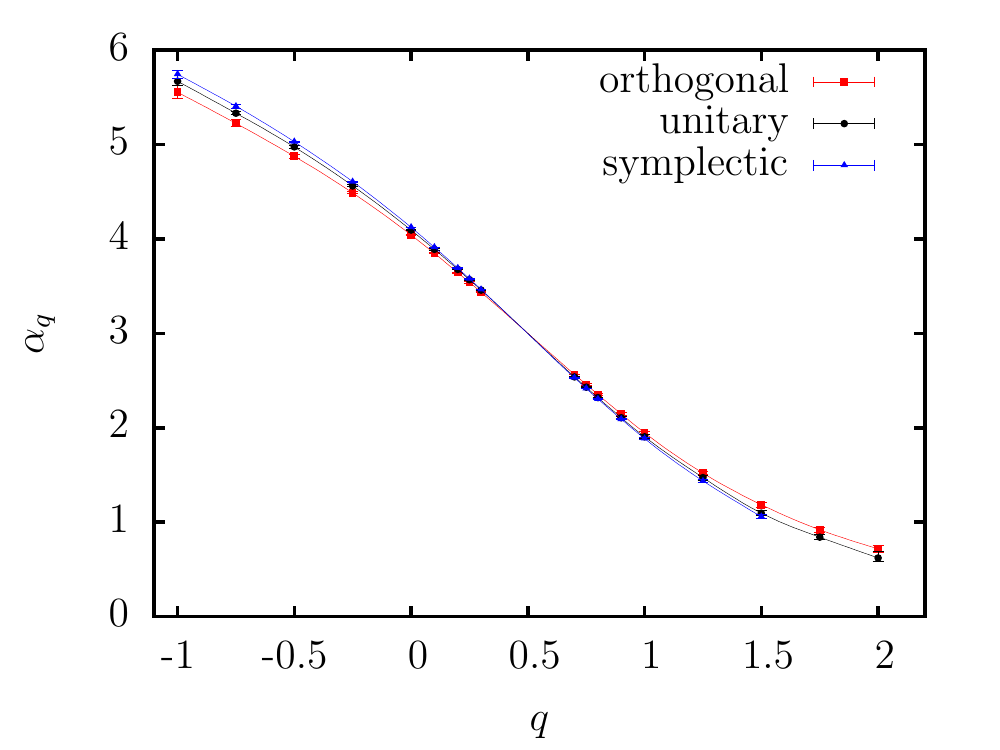} \\
  \includegraphics[type=pdf,ext=.pdf,read=.pdf,width=.25\textwidth]{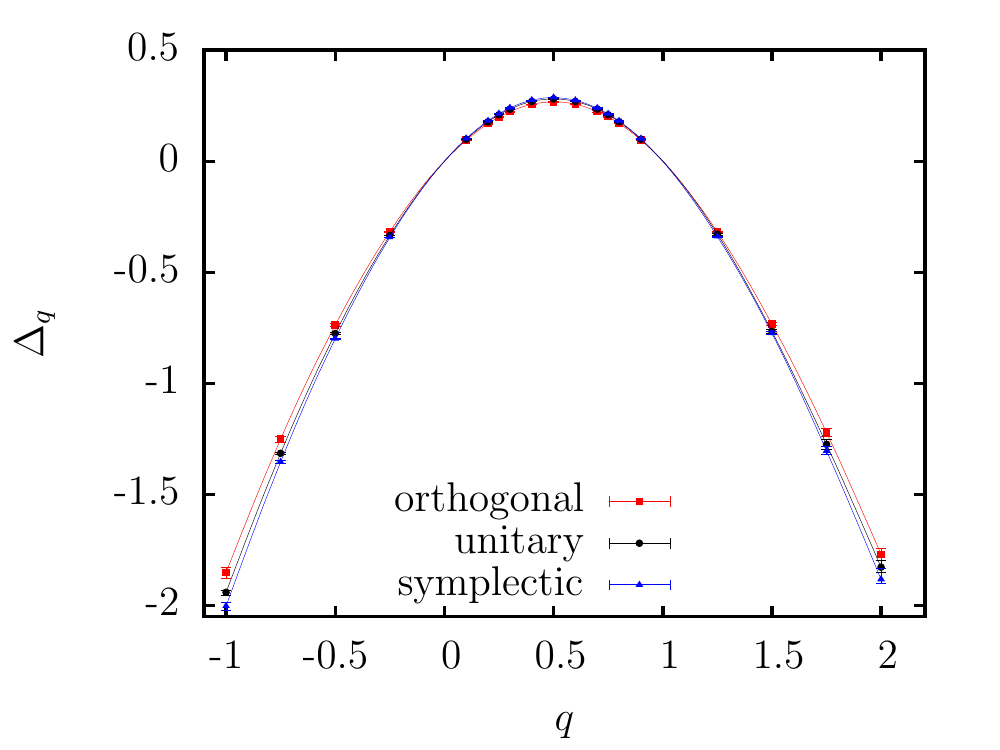} &
  \includegraphics[type=pdf,ext=.pdf,read=.pdf,width=.25\textwidth]{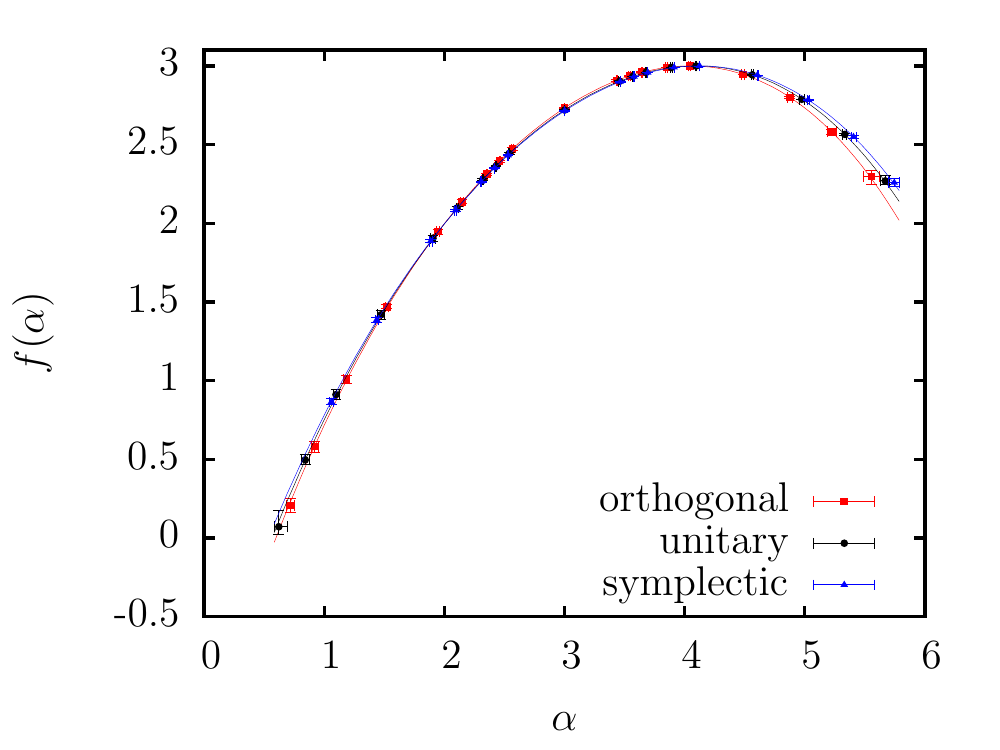} \\
  \end{tabular}
  \caption{MFEs of the Anderson models in the WD universality classes. Corresponding data are listed in 
  Tab.~\ref{tab:anderson_classes_MFEs_vary_lambda}.}
  \label{fig:anderson_classes_MFEs_vary_lambda}	
  \end{center}
\end{figure}

MFSS for varying $\lambda$ provided us the MFEs in all WD classes, which are listed in Tab.~\ref{tab:anderson_classes_MFEs_vary_lambda}, 
and depicted in Fig.~\ref{fig:anderson_classes_MFEs_vary_lambda}. For the orthogonal class one can find matching results with the listed MFE-s 
in Ref.~\onlinecite{Rodriguez11}. Since the precise values of the MFEs in three dimensions were determined first in Ref.~\onlinecite{Rodriguez11} 
for the orthogonal class only, the lack of reliable analytical and numerical results for the other symmetry classes makes our results more important. 
The most conspicuous thing in Fig.~\ref{fig:anderson_classes_MFEs_vary_lambda} is that curves for different symmetry classes are very close to 
each other, they are almost indistinguishable at the first sight. This shows that the broken time-reversal or spin rotational symmetry has a very small 
effect on the MFEs in three dimensions. Taking a closer look (or from Tab.~\ref{tab:anderson_classes_MFEs_vary_lambda}) one can see that the curve 
of $D_q$ and $\alpha_q$ are the steepest in the symplectic, the second steepest in the unitary, and the less steep in the orthogonal class. From 
Tab.~\ref{tab:anderson_classes_MFEs_vary_lambda} it is also clear that at most of the $q$ values there is a significant difference between the MFEs 
of different symmetry classes.

There are no critical states in the two dimensional orthogonal class~\cite{EversMirlin08}, but one can find values of $\alpha_0$ for the two dimensional 
unitary class (Integer Quantum Hall), $\alpha_{0\,2D}^{U}=2.2596\pm{0.0004}$~\cite{Evers08}, and symplectic class, 
$\alpha_{0\,2D}^{S}=2.172\pm{0.002}$~\cite{MildenbergerEvers07}. Comparing the difference between these exponents in two dimensions we get 
$\alpha_{0\,2D}^{U}-\alpha_{0\,2D}^{S}=0.0876\pm0.0024$, while our result for three dimensions is $\alpha_{0\,3D}^{U}-\alpha_{0\,3D}^{S}=-0.03\pm0.015$. 
There is about a factor of $3$ between the magnitude of these values, and even their sign is opposite, which shows very different effect of presence or 
absence of spin rotational symmetry in different dimensions.

\begin{figure}
  \begin {center}
  \begin{tabular}{c c}
  \includegraphics[type=pdf,ext=.pdf,read=.pdf,width=.25\textwidth]{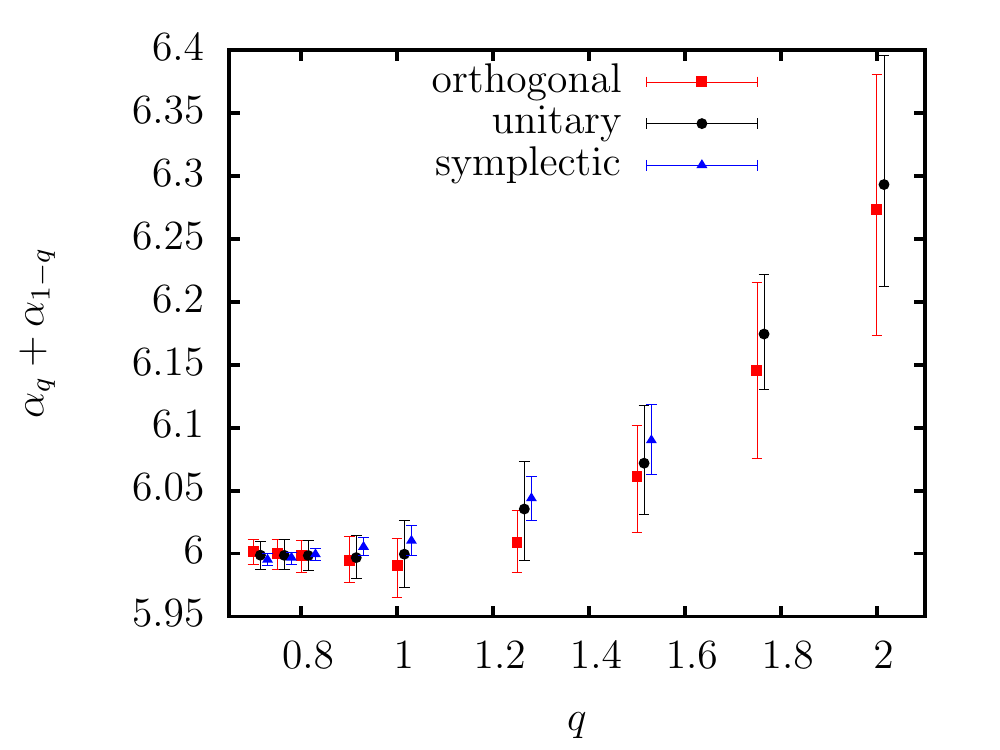} &
  \includegraphics[type=pdf,ext=.pdf,read=.pdf,width=.25\textwidth]{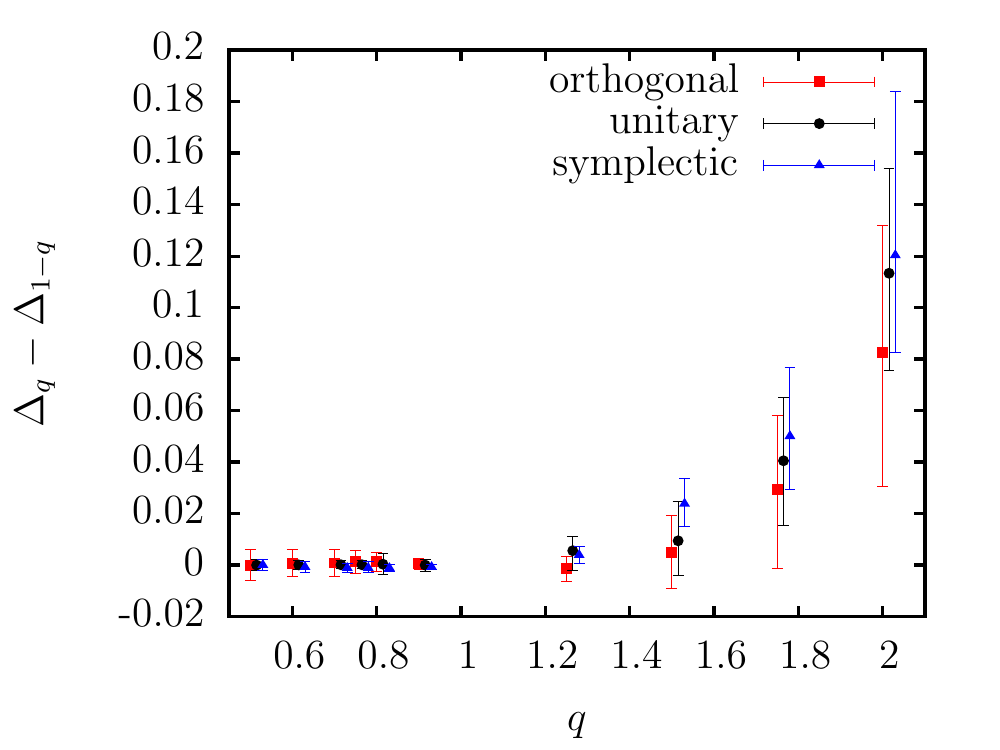} \\
  \end{tabular}
\caption{Test for symmetry relation Eq.~(\ref{eq:multifractals_Deltaalphasymmety}) in the WD symmetry classes. 
Points are shifted horizontally a little bit for better visualization. Only the range $q\geq 0.5$ is visible because 
expression $\alpha_q+\alpha_{1-q}$ ($\Delta_q-\Delta_{1-q}$) is symmetric (antisymmetric) for $q=0.5$.}
  \label{fig:anderson_classes_alphaDeltaqsymm_vary_lambda}	
  \end{center}
\end{figure}

We tested the symmetry relation Eq.~(\ref{eq:multifractals_Deltaalphasymmety}) for $\alpha_q$ and $\Delta_q$, 
the results are listed in Tab.~\ref{tab:anderson_classes_MFEs_vary_lambda} and depicted in 
Fig.~\ref{fig:anderson_classes_alphaDeltaqsymm_vary_lambda}. The symmetry relation is fulfilled in the 
range $-0.25\leq q\leq 1.25$ (in the {\it symplectic} class only for $-0.25\leq q\leq 1$), and small deviations are visible outside this interval. 
In this regime error bars are growing very large, coming mainly from the large errors of $\alpha_{q\geq 1.5}$ and $D_{q\geq 1.5}$. 
Similar effects were already seen for the critical parameters 
in Fig.~\ref{fig:anderson_classes_fitres_vary_lambda}. It is really hard to estimate the correct error bars in this 
large $q$ case, and the deviations from symmetry are small, therefore we believe that differences appear only 
because of slightly underestimated error bars of $\alpha_{q\geq 1.5}$ and $D_{q\geq 1.5}$. 
All in all we find numerical results basically matching with Eq.~(\ref{eq:multifractals_Deltaalphasymmety}).

\begin{figure}
  \begin {center}
  \begin{tabular}{c c}
  \includegraphics[type=pdf,ext=.pdf,read=.pdf,width=.25\textwidth]{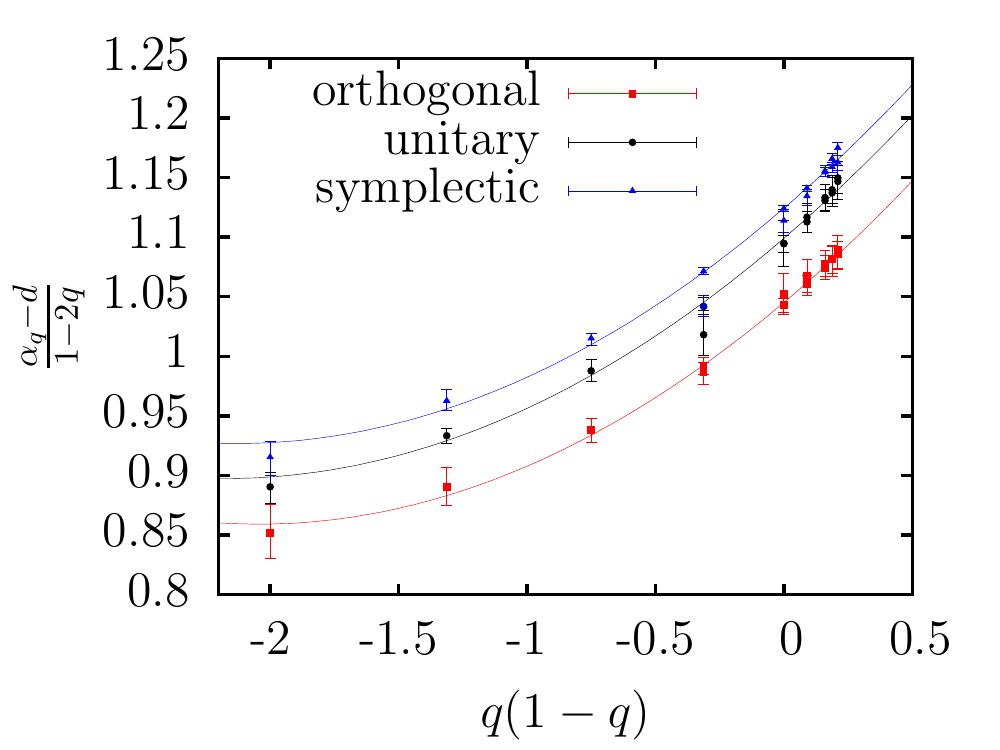} &
  \includegraphics[type=pdf,ext=.pdf,read=.pdf,width=.25\textwidth]{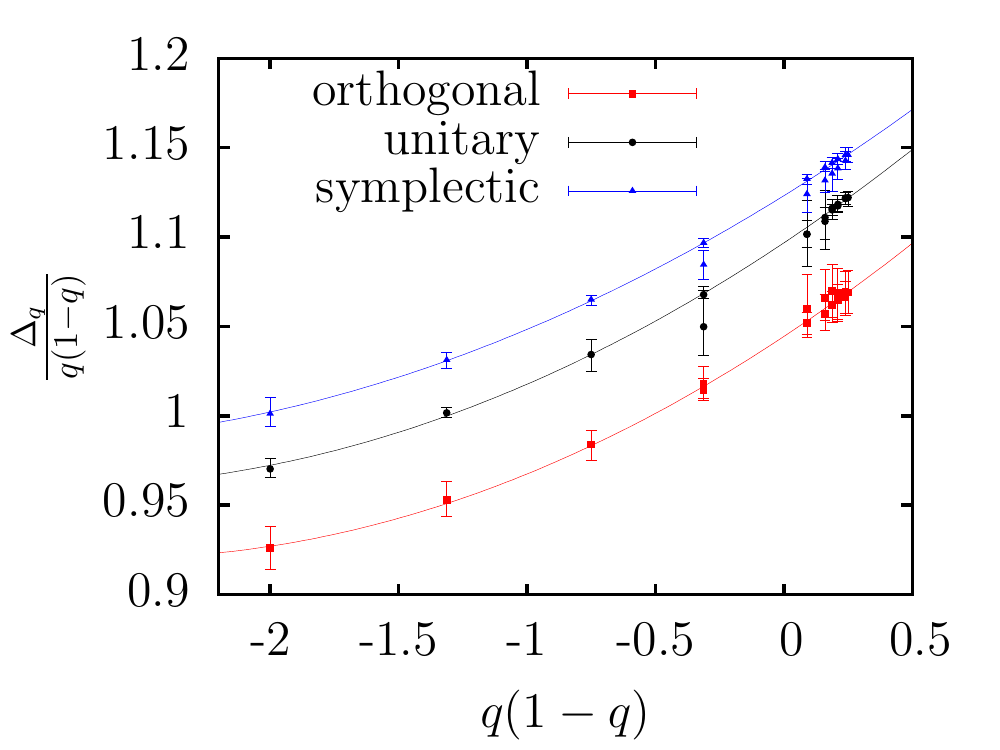} \\
  \end{tabular}
\caption{Dots and error bars are numerical values for the corresponding quantities, $\frac{\alpha(q)-d}{1-2q}$ 
and $\frac{\Delta(q)}{q(1-q)}$, for the WD symmetry classes. Lines are the best fits. Several points are shifted 
horizontally a bit for better viewing.}
  \label{fig:anderson_classes_alphaDelta_q1mq_vary_lambda}	
  \end{center}
\end{figure}

Assuming, that $\Delta_q$ is an analytic function of $q$, and using the symmetry relation, 
Eq.~(\ref{eq:multifractals_Deltaalphasymmety}), one can expand $\Delta_q$ in Taylor series around $q=\frac{1}{2}$:
\begin{eqnarray} 
\Delta_q &=& \sum_{k=0}^\infty c_k \left(q-\frac{1}{2}\right)^{2k}=
\sum_{k=0}^\infty c_k \left(q(q-1)+\frac{1}{4}\right)^{k}= \nonumber\\
&=& \sum_{k=0}^\infty c_k \sum_{i=0}^k \binom{k}{i}\left(q(q-1)\right)^{i}\left(\frac{1}{4}\right)^{k-i}=\nonumber\\
&=& \sum_{k=1}^\infty d_k \left(q(1-q)\right)^{k},
\label{eq:anderson_classes_Deltaexpansion_vary_lambda}\end{eqnarray}
where the condition $\Delta_0=\Delta_1=0$ enforced by the definition of $\Delta_q$ 
(see Eq.~(\ref{eq:multifractals_tau_D_Delta})) was used in the last step, leading to $k=1$ as the
lower bound for the summation. Similar expression can be derived for $\alpha_q$ by using the 
connection $\alpha_q=d+\frac{d}{dq}\Delta_q$ derived from 
Eqs.~(\ref{eq:multifractals_tau_alpha})--(\ref{eq:multifractals_tau_D_Delta}):
\begin{equation} 
\alpha_q = d+(1-2q)\sum_{k=1}^\infty a_k \left(q(1-q)\right)^{k-1},
\label{eq:anderson_classes_alphaexpansion_vary_lambda}
\end{equation}
where $a_k=kd_k$, and $a_1=d_1=\alpha_0-d$. One can obtain the $d_k$ and $a_k$ coefficients by fitting the expressions 
Eq.~(\ref{eq:anderson_classes_Deltaexpansion_vary_lambda})--(\ref{eq:anderson_classes_alphaexpansion_vary_lambda}). 
We used only the range $q\leq 1.25$, because beyond this regime error bars are 
growing extremely large, and there are small deviations from the symmetry relation 
Eq.(\ref{eq:multifractals_Deltaalphasymmety}) also. We plotted $\frac{\Delta(q)}{q(1-q)}$ and 
$\frac{\alpha(q)-d}{1-2q}$ in Fig.~\ref{fig:anderson_classes_alphaDelta_q1mq_vary_lambda} to make the presence of higher-order terms of the expansion visible.

We fit expressions 
Eq.~(\ref{eq:anderson_classes_Deltaexpansion_vary_lambda})--(\ref{eq:anderson_classes_alphaexpansion_vary_lambda})
up to third order in all cases, the resulting expansion coefficients are listed in 
Tab.~\ref{tab:anderson_classes_expcoeff_vary_lambda}. From the  data listed one can see that the 
expansion coefficients fulfill the relation $a_k=kd_k$. However $\alpha_q$ and $\Delta_q$ were obtained 
from the same wave-functions, they are results of completely independent fit-procedures. Therefore the fact, 
that they satisfy the equation  $a_k=kd_k$ further confirms our result for their value listed in 
Tab.~\ref{tab:anderson_classes_MFEs_vary_lambda} for $q\leq 1.25$ and shows the consistency of the MFSS.

\begin{table}
  \begin {center}
  \begin{tabular}{|c| c| c| c|}
  \hline
   & ort & uni & sym \\ \hline
  $d_1$ & $1.044\ (1.041..1.047)$ & $1.097\ (1.095..1.098)$ & $1.123\ (1.122..1.125)$ \\ \hline
  $d_2$ & $0.095\ (0.085..0.105)$ & $0.096\ (0.091..0.100)$ & $0.088\ (0.084..0.093)$ \\ \hline 
  $d_3$ & $0.018\ (0.011..0.025)$ & $0.017\ (0.014..0.020)$ & $0.014\ (0.010..0.017)$ \\ \hline
  $a_1$ & $1.045\ (1.042..1.048)$ & $1.099\ (1.096..1.102)$ & $1.124\ (1.123..1.126)$ \\ \hline
  $a_2$ & $0.182\ (0.168..0.195)$ & $0.185\ (0.174..0.197)$ & $0.185\ (0.179..0.191)$ \\ \hline
  $a_3$ & $0.044\ (0.035..0.053)$ & $0.043\ (0.035..0.050)$ & $0.044\ (0.038..0.049)$ \\ \hline
\end{tabular}
\caption{Expansion coefficients of 
Eqs.~(\ref{eq:anderson_classes_Deltaexpansion_vary_lambda})--(\ref{eq:anderson_classes_alphaexpansion_vary_lambda}) 
obtained by a fit depicted in Fig.~\ref{fig:anderson_classes_alphaDelta_q1mq_vary_lambda}.}
\label{tab:anderson_classes_expcoeff_vary_lambda}
\end{center}
\end{table}

As one would expect for expansion coefficients, $d_k$ and $a_k$ show decreasing behavior as 
$k$ grows. Only $d_1$ and $a_1$ are significantly different for the different symmetry classes, while 
$d_2,\ d_3,\ a_2$ and $a_3$ are the same within error bars. Their real value is probably different, 
but the relative error of the expansion coefficients naturally increases as $k$ grows, leading to 
indistinguishable values for the different symmetry classes for $k\geq 2$. 

Wegner computed analytically~\cite{Wegner87} the value of $\Delta_q$ with $\varepsilon$ expansion 
using nonlinear $\sigma$-model up to fourth-loop order for the orthogonal and the unitary symmetry class, 
resulting an expansion in dimensions $d=2+\varepsilon$ for $\varepsilon\ll 1$~\cite{EversMirlin08}:

\begin{align}
&\Delta_q^O = q(1-q)\varepsilon+\frac{\zeta(3)}{4}q(q-1)(q^2-q+1)\varepsilon^4+\mathcal{O}(\varepsilon^5)\nonumber\\
\label{eq:anderson_classes_Wegner_1} &= \left(\varepsilon-\frac{\zeta(3)}{4}\varepsilon^4\right)q(1-q)+\frac{\zeta(3)}{4}\varepsilon^4(q(1-q))^2+\mathcal{O}(\varepsilon^5)\\
\label{eq:anderson_classes_Wegner_2} &\Delta_q^U = \sqrt{\frac{\varepsilon}{2}}q(1-q)-\frac{3}{8}\zeta(3)\varepsilon^2(q(1-q))^2+\mathcal{O}(\varepsilon^{\frac{5}{2}})
\end{align}
Even though $\varepsilon\ll 1$ should hold, one can try to extrapolate to three-dimensions by inserting $\varepsilon=1$. 
This leads to $d_1^O\approx 0.699$, $d_2^O\approx 0.301$, $d_1^U\approx 0.707$ and $d_2^U\approx -0.451$. 
As one can see, these values are rather far from our numerical results, but this is not surprising for an 
$\varepsilon$-expansion at $\varepsilon=1$. These results capture well the tendency at least that $d_1^O$ is slightly smaller, than $d_1^U$. 
On the other hand it leads to $d_2^O$ and $d_2^U$ having opposite sign, which is highly inconsistent with our numerical results. It is interesting that the first-loop term, which is proportional to $\varepsilon$ and leads to 
parabolic $\Delta_q$, results in $d_1^O=1$ and $a_1^O=\alpha_0-d=1$, which are very close to our numerically measured values. In this sense parabolic 
approximation is better for the orthogonal class, as compared to the fourth-loop order approximation.
If higher-order terms were obtained, or if $\Delta_q$ were expanded by using another approach, our coefficients could provide relatively accurate values 
as compared with analytical results. 
\begin{table*}
	\begin {center}
	{
	\renewcommand{\arraystretch}{0.7}
	\begin{tabular}{|c| c| c| c| c| c| c|}
	\hline

{\scriptsize q} & {\scriptsize class} & $\scriptstyle{\alpha_q}$ & $\scriptstyle{D_q}$ & $\scriptstyle{f(\alpha_q)}$ & $\scriptstyle{\alpha_q+\alpha_{1-q}}$ & $\scriptstyle{\Delta_q-\Delta_{1-q}}$\\ \hline
\multirow{3}{*}{$\scriptstyle{-1}$} & {\scriptsize ort} & $\scriptstyle{5.555\ (5.490..5.626)}$ & $\scriptstyle{3.926\ (3.914..3.938)}$ & $\scriptstyle{2.297\ (2.338..2.250)}$ & $\scriptstyle{6.275\ (6.042..6.661)}$ & $\scriptstyle{-0.102\ (-0.218..0.000)}$ \\
& {\scriptsize uni} & $\scriptstyle{5.671\ (5.629..5.707)}$ & $\scriptstyle{3.970\ (3.966..3.976)}$ & $\scriptstyle{2.269\ (2.303..2.245)}$ & $\scriptstyle{6.331\ (6.215..6.444)}$ & $\scriptstyle{-0.130\ (-0.195..-0.062)}$ \\
& {\scriptsize sym} & $\scriptstyle{5.751\ (5.690..5.799)}$ & $\scriptstyle{4.001\ (3.994..4.010)}$ & $\scriptstyle{2.251\ (2.298..2.222)}$ & $\scriptstyle{6.379\ (6.197..6.584)}$ & $\scriptstyle{-0.134\ (-0.237..-0.063)}$ \\ \hline
 
\multirow{3}{*}{$\scriptstyle{-0.75}$} & {\scriptsize ort} & $\scriptstyle{5.225\ (5.187..5.267)}$ & $\scriptstyle{3.715\ (3.708..3.722)}$ & $\scriptstyle{2.582\ (2.599..2.564)}$ & $\scriptstyle{6.153\ (5.988..6.353)}$ & $\scriptstyle{-0.035\ (-0.094..0.032)}$ \\
& {\scriptsize uni} & $\scriptstyle{5.333\ (5.317..5.349)}$ & $\scriptstyle{3.751\ (3.749..3.754)}$ & $\scriptstyle{2.565\ (2.573..2.557)}$ & $\scriptstyle{6.176\ (6.131..6.239)}$ & $\scriptstyle{-0.062\ (-0.098..-0.025)}$ \\
& {\scriptsize sym} & $\scriptstyle{5.406\ (5.387..5.430)}$ & $\scriptstyle{3.773\ (3.770..3.777)}$ & $\scriptstyle{2.549\ (2.558..2.537)}$ & $\scriptstyle{6.221\ (6.113..6.349)}$ & $\scriptstyle{-0.060\ (-0.114..-0.023)}$ \\ \hline
 
\multirow{3}{*}{$\scriptstyle{-0.5}$} & {\scriptsize ort} & $\scriptstyle{4.876\ (4.856..4.896)}$ & $\scriptstyle{3.492\ (3.488..3.496)}$ & $\scriptstyle{2.800\ (2.803..2.796)}$ & $\scriptstyle{6.061\ (5.959..6.149)}$ & $\scriptstyle{-0.008\ (-0.045..0.025)}$ \\
& {\scriptsize uni} & $\scriptstyle{4.975\ (4.958..4.994)}$ & $\scriptstyle{3.517\ (3.512..3.521)}$ & $\scriptstyle{2.788\ (2.789..2.785)}$ & $\scriptstyle{6.103\ (6.000..6.167)}$ & $\scriptstyle{-0.009\ (-0.025..0.004)}$ \\
& {\scriptsize sym} & $\scriptstyle{5.030\ (5.019..5.039)}$ & $\scriptstyle{3.532\ (3.531..3.534)}$ & $\scriptstyle{2.784\ (2.787..2.781)}$ & $\scriptstyle{6.103\ (6.039..6.206)}$ & $\scriptstyle{-0.019\ (-0.041..-0.001)}$ \\ \hline
 
\multirow{3}{*}{$\scriptstyle{-0.25}$} & {\scriptsize ort} & $\scriptstyle{4.488\ (4.477..4.499)}$ & $\scriptstyle{3.254\ (3.252..3.255)}$ & $\scriptstyle{2.945\ (2.946..2.944)}$ & $\scriptstyle{6.016\ (5.951..6.094)}$ & $\scriptstyle{0.000\ (-0.012..0.010)}$ \\
& {\scriptsize uni} & $\scriptstyle{4.563\ (4.553..4.574)}$ & $\scriptstyle{3.267\ (3.266..3.268)}$ & $\scriptstyle{2.943\ (2.945..2.941)}$ & $\scriptstyle{6.037\ (5.998..6.081)}$ & $\scriptstyle{-0.006\ (-0.011..0.002)}$ \\
& {\scriptsize sym} & $\scriptstyle{4.607\ (4.603..4.611)}$ & $\scriptstyle{3.274\ (3.274..3.275)}$ & $\scriptstyle{2.941\ (2.941..2.941)}$ & $\scriptstyle{6.033\ (5.997..6.072)}$ & $\scriptstyle{-0.004\ (-0.011..0.003)}$ \\ \hline
 
\multirow{3}{*}{$\scriptstyle{0}$} & {\scriptsize ort} & $\scriptstyle{4.043\ (4.035..4.049)}$ & $\scriptstyle{3\ (3..3)}$ & $\scriptstyle{3\ (3..3)}$ & $\scriptstyle{5.991\ (5.965..6.012)}$ & $\scriptstyle{0\ (0..0)}$ \\
& {\scriptsize uni} & $\scriptstyle{4.094\ (4.087..4.101)}$ & $\scriptstyle{3\ (3..3)}$ & $\scriptstyle{3\ (3..3)}$ & $\scriptstyle{6.000\ (5.974..6.026)}$ & $\scriptstyle{0\ (0..0)}$ \\
& {\scriptsize sym} & $\scriptstyle{4.124\ (4.121..4.127)}$ & $\scriptstyle{3\ (3..3)}$ & $\scriptstyle{3\ (3..3)}$ & $\scriptstyle{6.010\ (5.999..6.023)}$ & $\scriptstyle{0\ (0..0)}$ \\ \hline
 
\multirow{3}{*}{$\scriptstyle{0.1}$} & {\scriptsize ort} & $\scriptstyle{3.849\ (3.843..3.855)}$ & $\scriptstyle{2.895\ (2.894..2.895)}$ & $\scriptstyle{2.990\ (2.989..2.991)}$ & $\scriptstyle{5.995\ (5.978..6.014)}$ & $\scriptstyle{-0.001\ (-0.003..0.002)}$ \\
& {\scriptsize uni} & $\scriptstyle{3.890\ (3.883..3.897)}$ & $\scriptstyle{2.890\ (2.889..2.891)}$ & $\scriptstyle{2.990\ (2.988..2.991)}$ & $\scriptstyle{5.997\ (5.981..6.014)}$ & $\scriptstyle{0.000\ (-0.002..0.002)}$ \\
& {\scriptsize sym} & $\scriptstyle{3.913\ (3.911..3.915)}$ & $\scriptstyle{2.887\ (2.886..2.887)}$ & $\scriptstyle{2.989\ (2.989..2.990)}$ & $\scriptstyle{6.005\ (5.998..6.013)}$ & $\scriptstyle{0.001\ (-0.000..0.002)}$ \\ \hline
 
\multirow{3}{*}{$\scriptstyle{0.2}$} & {\scriptsize ort} & $\scriptstyle{3.645\ (3.638..3.651)}$ & $\scriptstyle{2.789\ (2.786..2.790)}$ & $\scriptstyle{2.960\ (2.957..2.962)}$ & $\scriptstyle{5.998\ (5.985..6.011)}$ & $\scriptstyle{-0.001\ (-0.005..0.003)}$ \\
& {\scriptsize uni} & $\scriptstyle{3.678\ (3.673..3.684)}$ & $\scriptstyle{2.778\ (2.777..2.780)}$ & $\scriptstyle{2.958\ (2.956..2.961)}$ & $\scriptstyle{5.999\ (5.987..6.011)}$ & $\scriptstyle{-0.000\ (-0.004..0.004)}$ \\
& {\scriptsize sym} & $\scriptstyle{3.693\ (3.691..3.695)}$ & $\scriptstyle{2.772\ (2.772..2.773)}$ & $\scriptstyle{2.956\ (2.955..2.957)}$ & $\scriptstyle{5.999\ (5.995..6.004)}$ & $\scriptstyle{0.001\ (-0.000..0.003)}$ \\ \hline
 
\multirow{3}{*}{$\scriptstyle{0.25}$} & {\scriptsize ort} & $\scriptstyle{3.541\ (3.534..3.547)}$ & $\scriptstyle{2.734\ (2.733..2.737)}$ & $\scriptstyle{2.936\ (2.933..2.939)}$ & $\scriptstyle{6.000\ (5.987..6.012)}$ & $\scriptstyle{-0.001\ (-0.006..0.003)}$ \\
& {\scriptsize uni} & $\scriptstyle{3.569\ (3.563..3.575)}$ & $\scriptstyle{2.721\ (2.720..2.722)}$ & $\scriptstyle{2.933\ (2.931..2.935)}$ & $\scriptstyle{5.999\ (5.987..6.011)}$ & $\scriptstyle{-0.000\ (-0.002..0.001)}$ \\
& {\scriptsize sym} & $\scriptstyle{3.579\ (3.577..3.581)}$ & $\scriptstyle{2.715\ (2.714..2.715)}$ & $\scriptstyle{2.931\ (2.930..2.932)}$ & $\scriptstyle{5.997\ (5.992..6.001)}$ & $\scriptstyle{0.001\ (-0.001..0.003)}$ \\ \hline
 
\multirow{3}{*}{$\scriptstyle{0.3}$} & {\scriptsize ort} & $\scriptstyle{3.436\ (3.430..3.441)}$ & $\scriptstyle{2.681\ (2.678..2.684)}$ & $\scriptstyle{2.907\ (2.903..2.911)}$ & $\scriptstyle{6.001\ (5.991..6.012)}$ & $\scriptstyle{-0.001\ (-0.006..0.004)}$ \\
& {\scriptsize uni} & $\scriptstyle{3.459\ (3.453..3.464)}$ & $\scriptstyle{2.665\ (2.664..2.666)}$ & $\scriptstyle{2.903\ (2.900..2.905)}$ & $\scriptstyle{5.999\ (5.987..6.010)}$ & $\scriptstyle{-0.000\ (-0.002..0.001)}$ \\
& {\scriptsize sym} & $\scriptstyle{3.465\ (3.462..3.467)}$ & $\scriptstyle{2.657\ (2.656..2.658)}$ & $\scriptstyle{2.899\ (2.898..2.901)}$ & $\scriptstyle{5.995\ (5.991..6.000)}$ & $\scriptstyle{0.001\ (-0.001..0.003)}$ \\ \hline
 
\multirow{3}{*}{$\scriptstyle{0.4}$} & {\scriptsize ort} & $\scriptstyle{-}$ & $\scriptstyle{2.573\ (2.570..2.577)}$ & $\scriptstyle{-}$ & $\scriptstyle{-}$ & $\scriptstyle{-0.001\ (-0.006..0.004)}$ \\
& {\scriptsize uni} & $\scriptstyle{-}$ & $\scriptstyle{2.551\ (2.550..2.553)}$ & $\scriptstyle{-}$ & $\scriptstyle{-}$ & $\scriptstyle{-0.000\ (-0.002..0.002)}$ \\
& {\scriptsize sym} & $\scriptstyle{-}$ & $\scriptstyle{2.542\ (2.540..2.543)}$ & $\scriptstyle{-}$ & $\scriptstyle{-}$ & $\scriptstyle{0.001\ (-0.001..0.003)}$ \\ \hline
 
\multirow{3}{*}{$\scriptstyle{0.5}$} & {\scriptsize ort} & $\scriptstyle{3\ (3..3)}$ & $\scriptstyle{2.466\ (2.459..2.471)}$ & $\scriptstyle{2.733\ (2.730..2.736)}$ & $\scriptstyle{6\ (6..6)}$ & $\scriptstyle{0\ (0..0)}$ \\
& {\scriptsize uni} & $\scriptstyle{3\ (3..3)}$ & $\scriptstyle{2.439\ (2.437..2.441)}$ & $\scriptstyle{2.719\ (2.719..2.721)}$ & $\scriptstyle{6\ (6..6)}$ & $\scriptstyle{0\ (0..0)}$ \\
& {\scriptsize sym} & $\scriptstyle{3\ (3..3)}$ & $\scriptstyle{2.427\ (2.425..2.429)}$ & $\scriptstyle{2.714\ (2.712..2.715)}$ & $\scriptstyle{6\ (6..6)}$ & $\scriptstyle{0\ (0..0)}$ \\ \hline
 
\multirow{3}{*}{$\scriptstyle{0.6}$} & {\scriptsize ort} & $\scriptstyle{-}$ & $\scriptstyle{2.358\ (2.352..2.366)}$ & $\scriptstyle{-}$ & $\scriptstyle{-}$ & $\scriptstyle{0.001\ (-0.004..0.006)}$ \\
& {\scriptsize uni} & $\scriptstyle{-}$ & $\scriptstyle{2.327\ (2.325..2.329)}$ & $\scriptstyle{-}$ & $\scriptstyle{-}$ & $\scriptstyle{0.000\ (-0.002..0.002)}$ \\
& {\scriptsize sym} & $\scriptstyle{-}$ & $\scriptstyle{2.314\ (2.311..2.317)}$ & $\scriptstyle{-}$ & $\scriptstyle{-}$ & $\scriptstyle{-0.001\ (-0.003..0.001)}$ \\ \hline
 
\multirow{3}{*}{$\scriptstyle{0.7}$} & {\scriptsize ort} & $\scriptstyle{2.566\ (2.561..2.571)}$ & $\scriptstyle{2.252\ (2.242..2.263)}$ & $\scriptstyle{2.472\ (2.466..2.479)}$ & $\scriptstyle{6.001\ (5.991..6.012)}$ & $\scriptstyle{0.001\ (-0.004..0.006)}$ \\
& {\scriptsize uni} & $\scriptstyle{2.540\ (2.535..2.545)}$ & $\scriptstyle{2.217\ (2.214..2.220)}$ & $\scriptstyle{2.443\ (2.438..2.448)}$ & $\scriptstyle{5.999\ (5.987..6.010)}$ & $\scriptstyle{0.000\ (-0.001..0.002)}$ \\
& {\scriptsize sym} & $\scriptstyle{2.530\ (2.528..2.532)}$ & $\scriptstyle{2.203\ (2.199..2.207)}$ & $\scriptstyle{2.432\ (2.429..2.435)}$ & $\scriptstyle{5.995\ (5.991..6.000)}$ & $\scriptstyle{-0.001\ (-0.003..0.001)}$ \\ \hline
 
\multirow{3}{*}{$\scriptstyle{0.75}$} & {\scriptsize ort} & $\scriptstyle{2.459\ (2.454..2.465)}$ & $\scriptstyle{2.198\ (2.186..2.209)}$ & $\scriptstyle{2.394\ (2.387..2.401)}$ & $\scriptstyle{6.000\ (5.987..6.012)}$ & $\scriptstyle{0.001\ (-0.003..0.006)}$ \\
& {\scriptsize uni} & $\scriptstyle{2.430\ (2.424..2.436)}$ & $\scriptstyle{2.163\ (2.159..2.168)}$ & $\scriptstyle{2.363\ (2.358..2.369)}$ & $\scriptstyle{5.999\ (5.987..6.011)}$ & $\scriptstyle{0.000\ (-0.001..0.002)}$ \\
& {\scriptsize sym} & $\scriptstyle{2.417\ (2.415..2.419)}$ & $\scriptstyle{2.148\ (2.143..2.156)}$ & $\scriptstyle{2.350\ (2.347..2.353)}$ & $\scriptstyle{5.997\ (5.992..6.001)}$ & $\scriptstyle{-0.001\ (-0.003..0.001)}$ \\ \hline
 
\multirow{3}{*}{$\scriptstyle{0.8}$} & {\scriptsize ort} & $\scriptstyle{2.354\ (2.347..2.360)}$ & $\scriptstyle{2.147\ (2.135..2.157)}$ & $\scriptstyle{2.312\ (2.304..2.319)}$ & $\scriptstyle{5.998\ (5.985..6.011)}$ & $\scriptstyle{0.001\ (-0.003..0.005)}$ \\
& {\scriptsize uni} & $\scriptstyle{2.320\ (2.314..2.326)}$ & $\scriptstyle{2.111\ (2.099..2.125)}$ & $\scriptstyle{2.278\ (2.271..2.286)}$ & $\scriptstyle{5.999\ (5.987..6.011)}$ & $\scriptstyle{0.000\ (-0.004..0.004)}$ \\
& {\scriptsize sym} & $\scriptstyle{2.307\ (2.304..2.309)}$ & $\scriptstyle{2.095\ (2.090..2.100)}$ & $\scriptstyle{2.264\ (2.261..2.267)}$ & $\scriptstyle{5.999\ (5.995..6.004)}$ & $\scriptstyle{-0.001\ (-0.003..0.000)}$ \\ \hline
 
\multirow{3}{*}{$\scriptstyle{0.9}$} & {\scriptsize ort} & $\scriptstyle{2.146\ (2.135..2.159)}$ & $\scriptstyle{2.046\ (2.029..2.060)}$ & $\scriptstyle{2.136\ (2.124..2.149)}$ & $\scriptstyle{5.995\ (5.978..6.014)}$ & $\scriptstyle{0.001\ (-0.002..0.003)}$ \\
& {\scriptsize uni} & $\scriptstyle{2.107\ (2.097..2.117)}$ & $\scriptstyle{2.009\ (1.991..2.025)}$ & $\scriptstyle{2.097\ (2.087..2.108)}$ & $\scriptstyle{5.997\ (5.981..6.014)}$ & $\scriptstyle{-0.000\ (-0.002..0.002)}$ \\
& {\scriptsize sym} & $\scriptstyle{2.092\ (2.088..2.099)}$ & $\scriptstyle{1.988\ (1.981..1.997)}$ & $\scriptstyle{2.082\ (2.077..2.088)}$ & $\scriptstyle{6.005\ (5.998..6.013)}$ & $\scriptstyle{-0.001\ (-0.002..0.000)}$ \\ \hline
 
\multirow{3}{*}{$\scriptstyle{1}$} & {\scriptsize ort} & $\scriptstyle{1.948\ (1.930..1.963)}$ & $\scriptstyle{\alpha_1}$ & $\scriptstyle{\alpha_1}$ & $\scriptstyle{5.991\ (5.965..6.012)}$ & $\scriptstyle{0\ (0..0)}$ \\
& {\scriptsize uni} & $\scriptstyle{1.905\ (1.886..1.925)}$ & $\scriptstyle{\alpha_1}$ & $\scriptstyle{\alpha_1}$ & $\scriptstyle{6.000\ (5.974..6.026)}$ & $\scriptstyle{0\ (0..0)}$ \\
& {\scriptsize sym} & $\scriptstyle{1.886\ (1.877..1.896)}$ & $\scriptstyle{\alpha_1}$ & $\scriptstyle{\alpha_1}$ & $\scriptstyle{6.010\ (5.999..6.023)}$ & $\scriptstyle{0\ (0..0)}$ \\ \hline
 
\multirow{3}{*}{$\scriptstyle{1.25}$} & {\scriptsize ort} & $\scriptstyle{1.520\ (1.508..1.535)}$ & $\scriptstyle{1.727\ (1.715..1.738)}$ & $\scriptstyle{1.477\ (1.418..1.551)}$ & $\scriptstyle{6.009\ (5.985..6.034)}$ & $\scriptstyle{-0.001\ (-0.006..0.003)}$ \\
& {\scriptsize uni} & $\scriptstyle{1.473\ (1.442..1.499)}$ & $\scriptstyle{1.688\ (1.660..1.708)}$ & $\scriptstyle{1.422\ (1.391..1.457)}$ & $\scriptstyle{6.036\ (5.995..6.073)}$ & $\scriptstyle{0.006\ (-0.002..0.011)}$ \\
& {\scriptsize sym} & $\scriptstyle{1.437\ (1.424..1.450)}$ & $\scriptstyle{1.644\ (1.634..1.655)}$ & $\scriptstyle{1.371\ (1.338..1.409)}$ & $\scriptstyle{6.044\ (6.027..6.061)}$ & $\scriptstyle{0.004\ (0.001..0.007)}$ \\ \hline
 
\multirow{3}{*}{$\scriptstyle{1.5}$} & {\scriptsize ort} & $\scriptstyle{1.185\ (1.161..1.206)}$ & $\scriptstyle{1.534\ (1.518..1.550)}$ & $\scriptstyle{1.007\ (0.912..1.079)}$ & $\scriptstyle{6.061\ (6.017..6.102)}$ & $\scriptstyle{0.005\ (-0.009..0.019)}$ \\
& {\scriptsize uni} & $\scriptstyle{1.096\ (1.073..1.124)}$ & $\scriptstyle{1.468\ (1.453..1.483)}$ & $\scriptstyle{0.958\ (0.836..1.017)}$ & $\scriptstyle{6.072\ (6.031..6.118)}$ & $\scriptstyle{0.009\ (-0.004..0.025)}$ \\
& {\scriptsize sym} & $\scriptstyle{1.060\ (1.044..1.080)}$ & $\scriptstyle{1.450\ (1.437..1.465)}$ & $\scriptstyle{0.889\ (0.827..1.011)}$ & $\scriptstyle{6.090\ (6.063..6.118)}$ & $\scriptstyle{0.024\ (0.015..0.034)}$ \\ \hline
 
\multirow{3}{*}{$\scriptstyle{1.75}$} & {\scriptsize ort} & $\scriptstyle{0.920\ (0.889..0.949)}$ & $\scriptstyle{1.372\ (1.349..1.395)}$ & $\scriptstyle{0.590\ (0.422..0.818)}$ & $\scriptstyle{6.145\ (6.076..6.216)}$ & $\scriptstyle{0.029\ (-0.001..0.058)}$ \\
& {\scriptsize uni} & $\scriptstyle{0.841\ (0.814..0.873)}$ & $\scriptstyle{1.301\ (1.273..1.329)}$ & $\scriptstyle{0.479\ (0.459..0.529)}$ & $\scriptstyle{6.175\ (6.130..6.222)}$ & $\scriptstyle{0.041\ (0.015..0.065)}$ \\
& {\scriptsize sym} & $\scriptstyle{\text{no stability}}$ & $\scriptstyle{1.262\ (1.242..1.290)}$ & $\scriptstyle{}$ & $\scriptstyle{}$ & $\scriptstyle{0.050\ (0.030..0.077)}$ \\ \hline
 
\multirow{3}{*}{$\scriptstyle{2}$} & {\scriptsize ort} & $\scriptstyle{0.719\ (0.683..0.754)}$ & $\scriptstyle{1.231\ (1.203..1.256)}$ & $\scriptstyle{0.190\ (-0.068..0.727)}$ & $\scriptstyle{6.274\ (6.173..6.380)}$ & $\scriptstyle{0.083\ (0.031..0.132)}$ \\
& {\scriptsize uni} & $\scriptstyle{0.622\ (0.583..0.690)}$ & $\scriptstyle{1.173\ (1.147..1.205)}$ & $\scriptstyle{0.131\ (0.039..0.230)}$ & $\scriptstyle{6.293\ (6.212..6.396)}$ & $\scriptstyle{0.113\ (0.076..0.154)}$ \\
& {\scriptsize sym} & $\scriptstyle{\text{no stability}}$ & $\scriptstyle{1.118\ (1.099..1.167)}$ & $\scriptstyle{}$ & $\scriptstyle{}$ & $\scriptstyle{0.120\ (0.083..0.184)}$ \\ \hline
 
\end{tabular}
}
\caption{MFE $\alpha_q$, $D_q$ and $f(\alpha_q)$, and values for the corresponding symmetry relation 
Eq.~(\ref{eq:multifractals_Deltaalphasymmety}) obtained for the WD symmetry classes.}
\label{tab:anderson_classes_MFEs_vary_lambda}
\end{center}
\end{table*}

\section{Summary}
\label{sec:summary}
In this paper we examined the three-dimensional Anderson models belonging to the conventional WD symmetry classes with the help of 
multifractal finite-size scaling using two methods: a simpler method for fixed $\lambda$ leading to a single-variable scaling function, and a 
more complicated one for varying $\lambda$ resulting in a two-variable scaling function. Both methods confirmed the presence of multifractality in all 
three symmetry classes, and we obtained critical parameters listed in Tabs.~\ref{tab:anderson_classes_fitres_lambda01} and 
~\ref{tab:anderson_classes_fitres_vary_lambda_chosen} in aggreement with each other and with previous results known from the literature. The more 
complicated varying $\lambda$ method provided more precise values for the critical parameters, listed in 
Tab.~\ref{tab:anderson_classes_fitres_vary_lambda_chosen}, and significantly different critical exponents for the different WD symmetry classes.

Applying the method of varying $\lambda$ we also calculated the multifractal exponents, that basically fulfill the expected symmetry relation Eq.
(\ref{eq:multifractals_Deltaalphasymmety}), small deviations were detected for large $q$-values probably due to slightly underestimated error bars. 
in Fig.~\ref{fig:anderson_classes_MFEs_vary_lambda} one can see that the MFEs of different symmetry classes are very close to each other, but 
Fig.~\ref{fig:anderson_classes_alphaDelta_q1mq_vary_lambda} or Tab.~\ref{tab:anderson_classes_MFEs_vary_lambda} shows significant differences 
between them for most of the values of $q$. We compared the difference of $\alpha_0$ in the unitary and symplectic class to available results in two 
dimensions, and we found completely different relation between the two and three dimensional cases.
We expanded the MFEs in terms of the variable $q(1-q)$, and determined the expansion coefficients up to third order numerically. The expansion coefficients of 
Eq.~(\ref{eq:anderson_classes_Deltaexpansion_vary_lambda})--(\ref{eq:anderson_classes_alphaexpansion_vary_lambda}) fulfill the expected relation 
$a_k=kd_k$ giving a further confirmation for the validity of our results for the MFEs listed in Tab.~\ref{tab:anderson_classes_MFEs_vary_lambda}. 
We also compared the numerical results to available analytical estimates, 
and found in some cases similar, but in other cases opposite qualitative behavior for expansion coefficients for the orthogonal and the unitary classes. 
Nevertheless, we believe that the numerical precision of our results should be used as tests for future renormalization or other type of 
expansion approximations. Therefore our results await analytical comparison.

\begin{acknowledgments}
Financial support from OTKA under Grant No. K108676, and Alexander von Humboldt Foundation is gratefully acknowledged.
\end{acknowledgments}

\end{document}